\documentclass[11pt]{article}

\usepackage{amsmath, euscript, amssymb, amsfonts, dsfont, stmaryrd}

\usepackage[matrix,arrow,curve]{xy}
\usepackage{mathrsfs}

\newcommand{\mS}{{\mathscr{S}}}

\newcommand{\hstar}{\mathbin{\hat{\star}}}


\newcommand{\oR}{{\mathbb R}}

\newcommand{\oC}{{\mathbb C}}

\newcommand{\oZ}{{\mathbb Z}}

\newcommand{\on}{{\mathsf n}}
\newcommand{\om}{{\mathsf m}}
\newcommand{\ok}{{\mathsf k}}

\newcommand{\defeq}{\overset{\text{def}}{=}}

\begin{document}

\bigskip

\hfill FIAN-TD/2010-15

\bigskip

\thispagestyle{empty}

\baselineskip=15pt

\vspace{2cm}

\begin{center}
{\Large\bf  Moyal multiplier algebras of the test function spaces
of type S}

\vspace{1cm}

{\large\bf M.~A.~Soloviev}\footnote{E-mail: soloviev@lpi.ru}

\vspace{0.5cm}

 \centerline{\sl P.~N.~Lebedev Physical Institute}
 \centerline{\sl Russian Academy of Sciences}
 \centerline{\sl  Leninsky Prospect 53, Moscow 119991, Russia}

\vskip 3em

\end{center}

\begin{abstract}

The Gel'fand-Shilov spaces of type $S$ are considered as
topological algebras with respect to  the Moyal star product and
their corresponding  algebras of multipliers are defined and
investigated. In contrast to the well-studied  case of Schwartz's
space $S$, these multipliers are allowed to have nonpolynomial
growth or infinite order singularities. The Moyal multiplication
is thereby extended to certain classes of ultradistributions,
hyperfunctions, and  analytic functionals. The main theorem of the
paper characterizes those elements of the dual of a given test
function space that are the Moyal multipliers of this space. The
smallest  nontrivial Fourier-invariant space in the scale of
$S$-type spaces is shown to play a special role, because its
corresponding Moyal multiplier algebra contains the largest
algebra of functions for which the power series defining their
star products are absolutely convergent. Furthermore, it contains
analogous algebras associated with cone-shaped regions, which can
be used  to formulate a causality condition in quantum field
theory on noncommutative space-time.

\end{abstract}

\vspace{1cm}

\noindent PACS numbers: 11.10.Nx, 02.30.Sa,  03.65.Ca

\noindent MSC-2000 classes: 53D55, 81S30, 81T05, 46L52, 46F10,
46F15,  46N50

\vspace{0.5cm}

\newpage

\setcounter{page}{2}

\newpage

\section{Introduction and motivation}
This paper continues the work started in~\cite{I}, where it was
shown that, under the condition  $\beta\ge\alpha$, the test
function spaces  $S^\beta_\alpha$ introduced by
Gel'fand-Shilov~\cite{GS2} are algebras with respect to the
Weyl-Groenewold-Moyal star product, or Moyal product as it is more
commonly called now.  The spaces $S^\beta_\alpha$ are smaller than
the Schwartz space $S$ of all smooth functions of fast decrease
and are specified by additional restrictions on smoothness and
behavior at infinity. Accordingly, their dual spaces
$(S^\beta_\alpha)'$ are larger than the space $S'$ of tempered
distributions, and this scale of spaces provides a wider framework
for various applications. The Moyal multiplication is a basic
notion of the Wigner-Weyl phase-space representation of quantum
mechanics because it is just the composition rule for functions on
the classical phase space that corresponds, via the Weyl
transformation, to the product of operators acting on a Hilbert
space. (And then the functions are treated as symbols of their
corresponding operators~\cite{BS}.) This notion is also central to
noncommutative quantum field theory (see, e.g,~\cite{Sz} for a
review), whose intensive development over the last years has been
stimulated by the in-depth analysis~\cite{DFR} of the quantum
limitations on localization of events in space-time  and by the
study~\cite{SWitten} of the low energy limit of string theory. The
Moyal product is a noncommutative deformation of the usual
pointwise product, generated by a Poisson structure. In the
simplest case of functions on the space $\oR^d$ endowed with a
linear Poisson structure, it can be written as
\begin{multline}
(f\star_\theta
g)(x)=f(x)\exp\left\{\frac{i}{2}\,\overleftarrow{\partial_i}\,
\theta^{ij}\,\overrightarrow{\partial_j}\right\}g(x) =
\left.\sum_{n=0}^\infty\left(\frac{i}{2}\,\frac{\partial}{\partial
x^i}\,\theta^{ij}\frac{\partial}{\partial y^j}\right)^n
f(x)g(y)\right|_{\,x=y},
 \label{1.1}
\end{multline}
where $\theta^{\mu\nu}$ is a real constant antisymmetric $d\times
d$ matrix. This formula goes back to the celebrated works of
Groenewold~\cite{Gr} and Moyal~\cite{Moyal}. The right-hand side
of~\eqref{1.1} is usually understood as a formal power series in
the noncommutativity parameter $\theta$ and the validity of this
formula is rather restrictive, though it is ideally suited for
polynomial symbols. There is another representation for the Moyal
star product, which came into use later and is easily  derivable
from~\eqref{1.1} under favorable assumptions about  $f$ and $g$.
Namely, if these functions decrease faster than any inverse power
of $|x|$ and their Fourier transforms $\hat f$ and $\hat g$
decrease faster than the Gaussian function, then the standard
theorems of analysis justify the formal application of the
operator $\exp\left\{\frac{i}{2}\,\partial_{x^i}\,
\theta^{ij}\,\partial_{y^j}\right\}$ to the Fourier representation
of $f(x)g(y)$ and we can write
 \begin{gather}
(f\star_\theta g)(x)=\frac{1}{(2\pi)^{2d}}\iint  \hat f(p)\hat
g(q)\,e^{i(p+q) x-\tfrac{i}{2}p\theta q} dp\, dq=\notag\\
\frac{1}{(2\pi)^{d}}\int  f(x-\tfrac12\theta q) \hat g(q)\,e^{iqx}
dq=\frac{1}{(2\pi)^{d}}\int  \hat f(p) g(x+\tfrac12\theta
p)\,e^{ipx} dp =\label{1.2} \\
\frac{1}{(2\pi)^{d}}\iint  f(x-\tfrac12\theta q) g(x+y)\,e^{-iqy}
 dq\, dy =\frac{1}{(2\pi)^{d}}\iint f(x+y) g(x+\tfrac12\theta
p)\,e^{-ipy}  dp\, dy\notag.
\end{gather}
where  $p\, x=\sum_{j=1}^dp_jx^j$ and $\theta
p=\sum_{j=1}^d\theta^{ij} p_j$. Clearly, the last two integrals
are well defined for any integrable $f$ and $g$ if the matrix
$\theta$ is invertible, and we obtain
 \begin{equation}
(f\star_\theta g)(x)=\frac{1}{\pi^{d}\det\theta}\iint f(x+y)
g(x+z)\,e^{-2iy\theta^{-1}z} dy dz.
 \label{1.3}
\end{equation}
(Note that the determinant of each invertible antisymmetric matrix
is positive.) Neither of definitions~\eqref{1.1} and~\eqref{1.3}
is universal, but the second one can be naturally extended to a
larger class of symbols by using the methods of distribution
theory, which opens the way  to unification.  It is readily  seen
that the Schwartz space $S$ is an associative topological algebra
with respect to product~\eqref{1.3}. Antonets~\cite{A1,A2} was the
first to propose the extension of~\eqref{1.3} by duality to those
elements of $S'$ that are  Moyal multipliers of $S$.  Such an
extension has been investigated by Kammerer~\cite{Kamm},
Maillard~\cite{Mail} and, most thoroughly, by Gracia-Bondia and
V\'arilly~\cite{G-BV1,G-BV2}. Here we generalize this approach to
the spaces $S^\beta_\alpha$. From the above it is clear that
different forms of   Moyal star product can be obtained by
extension (depending on the problem under consideration)  from an
appropriate function space, whereon~\eqref{1.1} and~\eqref{1.3}
are equally well defined and interconvertible. Because of this we
use the same notation for them  and we will sometimes write
$\star$ instead of $\star_\theta$ when this cannot cause
confusion.  The relation between formulas~\eqref{1.1} and
\eqref{1.3} was also discussed at length in~\cite{E},
where~\eqref{1.1} was  systematically treated as an asymptotic
expansion of~\eqref{1.3}.

For  the role of Moyal analysis in  quantum field theory on
noncommutative space-time, we refer the reader to~\cite{G-BV3}
and~\cite{G}. The question of causality is crucial for the
physical interpretation of noncommutative field theory and, when
analyzing its causal structure, it should be taken into account
that the Moyal star product is inherently nonlocal. As stressed
in~\cite{Alv}, the framework of tempered distributions is
apparently too restrictive for the nonperturbative study of
general properties of noncommutative field theories. In~\cite{I},
an enlarged framework with the use of test function spaces
$S^\beta_\alpha$ was proposed for this purpose and
series~\eqref{1.1}  was shown to converge for all elements of
$S^\beta_\alpha$ if and only if $\beta<1/2$. A similar proposal
was made in~\cite{Ch}. An examination~\cite{Green,S08} of locality
violations in some noncommutative models shows that they fail to
obey the microcausality axiom of the standard  quantum field
theory~\cite{SW, BLOT}. Nevertheless, they obey a weaker causality
condition~\cite{S07,S10} formulated  in terms of analytic test
functions. As argued elsewhere~\cite{S06}, this condition is
sufficient to ensure  such fundamental physical properties  as the
spin-statistics relation and the existence of CPT-symmetry. The
Fourier invariant spaces $S^\beta_\beta$ were used by Fischer and
Szabo~\cite{FS1,FS2} in their analysis of the renormalization
properties of scalar field theories on noncommutative Minkowski
space, but as shown by Zahn~\cite{Zahn}, this issue requires a
more careful study. Because of all the above, it is desirable to
construct and investigate the Moyal multiplier algebras of the
spaces of type $S$. We will show that for $\alpha=\beta$, these
algebras contain all polynomials and all distributions of compact
support as does the Moyal multiplier algebra of $S$. But in
contrast to the latter, they also contain a large class of
functions with non-polynomial growth and with non-tempered
singularities. In this paper, we focus on the extension of the
Weyl symbolic calculus  of itself. The results have direct
applications to deformation quantization and to field theory on
noncommutative spaces, but this applications will be considered in
a subsequent work.

The paper is organized as follows. In Sec.~II, we recall the
definition and  main properties of the spaces $S^\beta_\alpha$ and
their associated algebras of pointwise multipliers and of
convolution multipliers. The aim of this paper is to investigate a
noncommutative deformation of these algebras, but the basic
construction is presented in a more general setting, for an
arbitrary test function space $E\subset S$ allowing such
deformation. In Sec.~III, we define by duality the algebras
${\mathcal M}_{\theta,L}(E)$ and ${\mathcal M}_{\theta,R}(E)$ of
left and  right Moyal multipliers of $E$. In Sec.~IV, we show that
in the case of spaces $S^\beta_\alpha$, $\beta\ge\alpha$, every
Moyal multiplier can be approximated by test functions in the
operator topology. This result gives an alternative way of
defining the algebras ${\mathcal M}_{\theta,L}(S^\beta_\alpha)$
and ${\mathcal M}_{\theta,R}(S^\beta_\alpha)$, via a completion
procedure applied to the  sets of operators of the left and right
star multiplication by elements of $S^\beta_\alpha$. Moreover, the
approximation theorem lets us prove that the intersection
${\mathcal M}_\theta(S^\beta_\alpha)$ of these two algebras is
also an algebra. In the same section, we show that ${\mathcal
M}_\theta(S^\beta_\alpha)$ equipped with a natural topology acts
continuously on  the dual space $(S^\beta_\alpha)'$, i.e., this
space has the structure of an ${\mathcal
M}_\theta(S^\beta_\alpha)$-bimodule.  In Sec.~V, we characterize
the smoothness properties and the behavior at infinity of the
Fourier transforms of the star products $f\star_\theta u$ and
$u\star_\theta f$, where $f\in S^\beta_\alpha$ and $u\in
(S^\beta_\alpha)'$. In particular, we prove that for any matrix
$\theta$, the Fourier transforms of these products are pointwise
multipliers of $S_\beta^\alpha=\widehat{S^\beta_\alpha}$. Making
use of this result, we prove in Sec.~VI that the Moyal multiplier
algebra of the Fourier-invariant space $S^\beta_\beta$ contains
all elements of $(S^\beta_\beta)'$ decreasing  sufficiently fast
at infinity and, moreover, contains their Fourier transforms. In
Sec.~VII, we discuss a special role of the space $S^{1/2}_{1/2}$
and show that ${\mathcal M}_\theta(S^{1/2}_{1/2})$ contains the
largest algebra with the property that the Moyal power
series~\eqref{1.1} converges absolutely for all its elements.
Furthermore, we demonstrate that ${\mathcal
M}_\theta(S^{1/2}_{1/2})$ has a family of subalgebras associated
naturally with cone-shaped regions, which can be used for
formulating causality in a rigorous development of quantum field
theory on noncommutative space-time. Sec.~VIII is devoted to
concluding remarks. Appendix presents the proof of a lemma on  the
properties of the Fourier transforms of functions in
$S^\beta_\alpha$, which is used in deriving the main results.

\section{The pointwise multipliers of the spaces of type S}

We recall that the space $S^\beta_\alpha(\oR^d)$ with indices
$\alpha\ge0$, $\beta\ge 0$ consists of all infinitely
differentiable functions on $\oR^d$ satisfying the inequalities
\begin{equation}
|\partial^\on f(x)|\le C B^{|\on|}
\on^{\beta\on}e^{-|x/A|^{1/\alpha}} ,
 \label{2.1}
\end{equation}
where $C$, $A$, and $B$ are  constants depending on $f$ and
$\on=(n_1,\dots,n_d)$ is an arbitrary $d$-tuple of nonnegative
integers. From here on we use the standard multiindex notation:
$|\on|=n_1+\dots+n_d$, $\on^{\beta\on}=n_1^{\beta n_1}\dots
n_d^{\beta n_d}$, and $\partial^\on =\partial^{n_1}_{x_1}\dots
\partial^{n_d}_{x_d}$.  The above definition is independent of the
choice of the norm in $\oR^d$  because all these norms are
equivalent, but  the uniform norm $|x|=\max_{1\le j\le d}|x^j|$ is
most convenient for use below.  We write $S^\beta_\alpha$ instead
of $S^\beta_\alpha(\oR^d)$ when there is no risk of confusion. By
$S^{\beta, B}_{\alpha, A}$ we denote the set of functions
satisfying~\eqref{2.1} with fixed $A$ and $B$   and equip it with
the  norm
\begin{equation}
\|f \|_{A, B}=\sup_{x,\on}\left|\, \frac{\partial^\on
f(x)}{B^{|\on|}\on^{\on\beta}}\,e^{|x/A|^{1/\alpha}}\right|,
 \label{2.2}
\end{equation}
which turns $S^{\beta, B}_{\alpha, A}$ into a Banach space.
Accordingly, $S^\beta_\alpha$ is equipped with the inductive limit
topology determined by the canonical embeddings  $S^{\beta,
B}_{\alpha, A}\to S^\beta_\alpha$, hence it is a barrelled space.
The space $S^\beta_\alpha$ is nontrivial if and only if
$\alpha+\beta >1$ or $\alpha+\beta =1$ but  $\alpha$ and $\beta$
are nonzero. In what follows, we assume that this condition is
fulfilled. Every nontrivial space $S^\beta_\alpha$ is dense in
$S$. As shown in~\cite{GS2}, the natural maps $S^{\beta,
B}_{\alpha, A}\to S^{\beta, B'}_{\alpha, A'}$, where $A'>A$ and
$B'>B$,  are compact. It follows that $S^\beta_\alpha$ are DFS
spaces (dual Fr\'echet-Schwartz spaces). We refer the reader
to~\cite{MV} for the definition and properties of FS and DFS
spaces.  Furthermore, they are nuclear~\cite{M} and an analog of
the Schwartz kernel theorem holds for them. The pointwise
multiplication $(f,g)\to f\cdot g$ is well defined for elements of
$S^\beta_\alpha$ and is separately continuous~\cite{GS2}. Since
$S^\beta_\alpha$ is a DFS space, the separate  continuity implies
continuity, see, e.g.,~\cite{K}, Sec.~44.2, where the
corresponding theorem is proved even for a larger class of spaces.
Therefore every space $S^\beta_\alpha$ is a topological algebra
with respect to the pointwise product and with respect to the
(ordinary) convolution product. The Fourier transform is a
topological isomorphism between $S^\beta_\alpha$ and
$S_\beta^\alpha$ and converts one of these  products to the other.

The space of pointwise multipliers of  $S^\beta_\alpha$ has been
completely characterized by Palamodov~\cite{P}. Let $E$ be a
topological vector space contained in the Schwartz space $S$ and
let $\mathcal L(E)$ be the algebra of continuous linear operators
on $E$, equipped with the topology of uniform convergence on
bounded subsets of $E$. By Palamodov's definition, the space
$M(E)$ of multipliers of $E$ is the closure  in $\mathcal L(E)$ of
the set of all operators of multiplication by elements of $S$.
Accordingly, $M(E)$ is equipped with the topology induced by that
of $\mathcal L(E)$. The space $C(E)$ of convolution multipliers is
defined in a similar manner. Clearly, these spaces are subalgebras
of the operator algebra $\mathcal L(E)$. Let $E^{\beta,B}_{\alpha,
A}(\oR^d)$ be the space of all smooth functions $u$ on $\oR^d$
with the property that
\begin{equation}
\|u \|_{-A, B}=\sup_{x,\on}\left|\, \frac{\partial^\on
u(x)}{B^{|\on|}\on^{\on\beta}}\,e^{-|x/A|^{1/\alpha}}\right|<\infty.
 \label{2.4}
\end{equation}

{\bf Theorem~1\,}(Palamodov): {\it The spaces of pointwise
multipliers and convolution multipliers of $S^\beta_\alpha$ can be
presented  as follows:
\begin{gather}
M(S^\beta_\alpha)=\projlim_{A\to\infty}\injlim_{B\to\infty}
E^{\beta,B}_{\alpha,A}, \label{2.5}\\
 C(S^\beta_\alpha)=\projlim_{B\to\infty}\injlim_{A\to\infty}
(E^{\beta,B}_{\alpha,A})',
 \label{2.5*}
 \end{gather}
where  $(E^{\beta,B}_{\alpha,A})'$ is the dual of
$E^{\beta,B}_{\alpha,A}$, equipped with the strong topology. Both
these spaces are complete, nuclear, and semireflexive. The Fourier
transform is an isomorphism between $M(S^\beta_\alpha)$ and
$C(S^\alpha_\beta)$.}

The above-listed topological properties of the spaces~\eqref{2.5}
and \eqref{2.5*} follows from the corresponding  properties of
$S^\beta_\alpha$. In~\cite{P}, it was noted that these spaces are
quasicomplete. But if   $E$ is a complete DF space, then $\mathcal
L(E)$ is complete~\cite{K} and so is every closed subspace of
$\mathcal L(E)$. If  in addition $E$ is nuclear, than $\mathcal
L(E)$ is also nuclear~\cite{Sch} and hence semireflexive, and
these properties are also inherited by closed subspaces.

Since spaces~\eqref{2.5} and \eqref{2.5*} are embedded into
$(S^\beta_\alpha)'$, the Fourier transforms of their elements are
defined in the ordinary way: $\langle \hat u,\hat
f\rangle=(2\pi)^d\langle u,\check{f}\rangle$, where $f\in
S^\beta_\alpha$ and $\check{f}(x)=f(-x)$. The space
$S^\beta_\alpha$ is obviously contained in  either of these two
spaces and   is dense in both of them by Theorem~1. Hence their
duals $M'(S^\beta_\alpha)$ and $C'(S^\beta_\alpha)$ can also be
identified with linear subspaces of $(S^\beta_\alpha)'$, which are
described by the next theorem.

{\bf Theorem~2:} {\it The duals of $M(S^\beta_\alpha)$ and
$C(S^\beta_\alpha)$ have, respectively, the form
\begin{align}
M'(S_\alpha^\beta)&=\bigcup_A\bigcap_B(E_{\alpha,A}^{\beta,B})'
\label{2.6}\\
\intertext{and}
C'(S_\alpha^\beta)&=\bigcup\limits_B\bigcap\limits_A
E_{\alpha,A}^{\beta,B}.
 \label{2.7}
 \end{align}

Proof.} These formulas follow from the well known duality
relations~\cite{Sch} between projective and inductive limits. The
only subtlety is that  the projective limit is assumed to be in
reduced form in these relations. Let $\mathcal
E^{\beta,B}_{\alpha, A}\defeq \projlim_{\epsilon\to
0}E^{\beta,B+\epsilon}_{\alpha, A-\epsilon}$. A reasoning similar
to that used by Gel'fand and Shilov for $\projlim_{\epsilon\to
0}S^{\beta,B+\epsilon}_{\alpha, A+\epsilon}$  shows that the
natural maps $E^{\beta, B}_{\alpha, A}\to E^{\beta,
B+\epsilon}_{\alpha, A-\epsilon}$ are compact. Hence the spaces
$\mathcal E^{\beta,B}_{\alpha, A}$ are perfect, i.e., are FS
spaces (Fr\'echet-Schwartz spaces) in the modern terminology. In
particular, they are Montel spaces. For any
$A>A'>A^{\prime\prime}$ and $B<B'$, we have the commutative
diagram
  \begin{equation}
\xymatrix{
 E^{\beta,B}_{\alpha, A}\ar[r]\ar[dr]&\mathcal E^{\beta,B}_{\alpha,
 A'}\ar[d]\ar[r]&\mathcal E^{\beta,B'}_{\alpha, A^{\prime\prime}}\\
&E^{\beta,B'}_{\alpha, A'}\ar[ur]&
 }
 \label{diagramm}
  \end{equation}
where all arrows are natural embeddings. It follows that in
definition~\eqref{2.5}, the spaces $E^{\beta,B}_{\alpha, A}$ can
be replaced with $\mathcal E^{\beta,B}_{\alpha, A}$, leaving the
limit space unchanged.  We claim that $S_\alpha^\beta$ is dense in
$\injlim_B\mathcal E^{\beta,B}_{\alpha, A}$ for each $A>0$. Let
$f\in \mathcal E^{\beta,B}_{\alpha, A}$ and $2A_1<A$. Let $B_1$ be
large enough for  $S^{\beta,B_1}_{\alpha,A_1}(\oR^d)$ to be
nontrivial. We choose a function $e\in
S^{\beta,B_1}_{\alpha,A_1}(\oR^d)$ with the property $\int
e(\xi)d\xi=1$ and set $e_\nu(x)=\int_{|\xi|<\nu} e(x-\xi)d\xi$,
$\nu=1,2,\dots$. Then $e_\nu(x)\to1$ at every point $x$. Using
Leibniz's formula and the inequality
\begin{equation}-|x-\xi|^{1/\alpha}\le
-|x/2|^{1/\alpha}+|\xi|^{1/\alpha},
 \label{2.8}
\end{equation}
we obtain
\begin{multline}
|\partial^\on (f e_\nu)(x)|\le\\ C_\epsilon \sum_\om
\binom{\on}\om (B+\epsilon)^{|\om|}B_1^{|\on-\om|}
\om^{\beta\om}(\on-\om)^{\beta(\on-\om)}e^{|x/(A-\epsilon)|^{1/\alpha}}
\int_{|\xi|<\nu} e^{-|(x-\xi)/A_1|^{1/\alpha}}d\xi\le\\
C'_{\nu,\epsilon}(B+B_1+\epsilon)^{|\on|}
\on^{\alpha\on}e^{|x/(A-\epsilon)|^{1/\alpha}-|x/(2A_1)|^{1/\alpha}},
 \label{2.9}
\end{multline}
where $\epsilon>0$ can be taken arbitrarily small. Therefore, $f
e_\nu\in S_\alpha^\beta$.  On the other hand,
\begin{equation}
|\partial^\on (f e_\nu)(x)|\le C_\epsilon(B+B_1+\epsilon)^{|\on|}
\on^{\alpha\on}e^{|x/(A-\epsilon)|^{1/\alpha}}\int_{\oR^d}
e^{-|\xi/A_1|^{1/\alpha}}d\xi
 \label{2.10}
\end{equation}
and  the sequence $f e_\nu$ is hence bounded in  $\mathcal
E_{\alpha,A}^{\beta, B+B_1}$. Because it is a Montel space and its
topology is stronger than that of pointwise convergence, we
conclude that $f e_\nu\to f$ in $\mathcal E_{\alpha,A}^{\beta,
B+B_1}$.  This proves our claim. As a consequence,
$M(S_\alpha^\beta)$ is dense in $\injlim_B\mathcal
E^{\beta,B}_{\alpha, A}$ for any $A$, so the projective limit
$\projlim_A\{\injlim_B\mathcal E^{\beta,B}_{\alpha, A}\}$ is
reduced. Therefore, for every  $v\in M'(S_\alpha^\beta)$, there is
an $A$ such that $v$ has a unique continuous extension to
$\injlim_B\mathcal E_{\alpha,A}^{\beta, B}$ and thereby to each
space $\mathcal E_{\alpha,A}^{\beta, B}$, $B>0$. This is
equivalent to saying that for some $A$, the functional $v$ extends
continuously to each space $E^{\beta,B}_{\alpha, A}$, $B>0$. Thus,
formula~\eqref{2.6} is proved.

From the commutativity of diagram~\eqref{diagramm}, it follows
that the corresponding diagram for dual spaces is also
commutative, which shows that in~\eqref{2.5*} the spaces
$(E^{\beta,B}_{\alpha, A})'$ can be replaced by the DFS spaces
$(\mathcal E^{\beta,B}_{\alpha, A})'= \injlim_{\epsilon\to
0}(E^{\beta,B+\epsilon}_{\alpha, A-\epsilon})'$. By the
Hahn-Banach theorem, the functionals $\delta(x-\xi)$, where $\xi$
ranges over $\oR^d$, form a total set in every  $(\mathcal
E_{\alpha,A}^{\beta, B})'$ because any FS space is reflexive and
$(\mathcal E_{\alpha,A}^{\beta, B})^{\prime\prime}=\mathcal
E_{\alpha,A}^{\beta, B}$. The space $C(S_\alpha^\beta)$ contains
all these functionals and hence is dense in every $(\mathcal
E_{\alpha,A}^{\beta, B})'$, i.e., the projective limit
$\projlim_B\{\injlim_A(\mathcal E^{\beta,B}_{\alpha, A})'\}$ is
reduced. Therefore, for every $u\in C'(S_\alpha^\beta)$, there is
$B$ such that $u$ has a unique continuous extension to
$\injlim_A(\mathcal E_{\alpha,A}^{\beta, B})'$ and so belongs to
each of the spaces $\mathcal E_{\alpha,A}^{\beta, B}$, $A>0$. This
amounts to saying that for some $B$, the functional $u$ belongs to
each of the spaces $E^{\beta,B}_{\alpha, A}$, $A>0$.

\section{ Extension of the Moyal product by duality}

Let $E$ be a locally convex function space embedded  densely and
continuously  into the Schwartz space $S$. Then we have the
sequence of natural continuous injections
\begin{equation}
E\to S\to S'\to E'.
 \label{3*}
 \end{equation}
The third map in~\eqref{3*}, being the transpose of the first one,
is continuous for the strong as well as for the weak topology on
$E'$ (see~\cite{Sch}, Sec.~IV.7.4) and has a weakly dense image.
For $u\in E'$, we write $\langle u,f\rangle$ for the value of the
functional $u$ evaluated at $f\in E$. If $E$ is a topological
algebra under the Moyal multiplication, then the products $u\star
f$ and $f\star u$ can be defined by
\begin{equation}
\langle u\star f, g \rangle=\langle u,f\star g\rangle,\quad
\langle f\star u,g\rangle=\langle u, g\star f\rangle, \qquad g\in
E,
 \label{3.1}
\end{equation}
in complete analogy with the case $E=S$ studied in~[7]--[13].
Since the expressions on the right-hand side are linear and
continuous in $g$, these products are well defined as elements of
$E'$. From~\eqref{1.2}, it follows that
 \begin{equation}
\int (f\star g)(x)\,dx=\int f(x) g(x)\,dx,\qquad \text{for any
$f,g\in E$}.\label{3.*}
\end{equation}
This simple but important relation called the tracial property
implies that the products \eqref{3.1} are extensions of the
initial $\star$-multiplication on $E$. Indeed, using~\eqref{3.*}
and the associativity of the algebra  $(E,\star$), we obtain that
\begin{equation}
\langle h\star f, g\rangle=\langle h,f\star g\rangle=\langle
f,g\star h\rangle=\int(h\star g\star f)(x) dx, \quad \text{for all
$h,f,g\in E$}.
 \label{3.2}
\end{equation}
For every fixed $f$, the maps $u\to u\star f$ and $u\to f\star u$
of $E'$  into itself are continuous  because they are the
transposes of the continuous maps $g\to f\star g$ and $g\to g\star
f$. Since $E$ is dense in  $E'$, there are no other continuous
extensions of the $\star$-multiplication to the case where one of
factors belongs to $E'$. For every fixed $u\in E'$, the maps $f\to
f\star u$ and $f\to u\star f$ from $E$ into $E'$ are also
continuous. Consider for instance  the first of them. For any
$\epsilon>0$, we can find a  neighborhood $W$ of the origin in $E$
such that $u$ is bounded by $\epsilon$ on $W$. Because the map
$(f,g)\to f\star g$ is jointly continuous, there are neighborhoods
$U$ and $V$ in $E$ such that $f\star g\in W$ for all $f\in U$ and
all $g\in V$. For any bounded set $Q\subset E$, there is
$\delta>0$ such that $\delta Q\subset V$, hence $\sup_{g\in
Q}|\langle u, f\star g\rangle|\le \epsilon$ for any $f\in\delta
U$, which proves the statement. From~\eqref{3.1} and the
associativity of the $\star$-multiplication in $E$, it immediately
follows that
\begin{equation}
(u\star f)\star h=u\star(f\star h),\quad h\star(f\star u)=(h\star
f)\star u\quad\text{for all $u\in E'$, $f,h\in E$}.
 \label{3.3}
\end{equation}
This means that $E'$ has the structure of a (nonunital) bimodule
over the ring  $(E,\star)$.

Now we introduce the spaces of left and right $\star$-multipliers
of $E$:
\begin{gather}
{\mathcal M}_{\theta,L}(E)\defeq \{u\in E'\colon u\star_\theta
f\in E,\quad\text{for all  $f\in E$}\},\notag \\ {\mathcal
M}_{\theta,R}(E)\defeq \{u\in E'\colon  f\star_\theta u\in
E,\quad\text{for all $f\in E$}\}. \notag
\end{gather}
The linear maps $f\to u\star f$ and $f\to  f\star u$ of $E$ into
itself have closed graphs.  Indeed, if $f_\nu\to f$ and $u\star
f_\nu\to h$, then for any $g\in E$, we have
\begin{equation}
\langle h,g\rangle=\lim\nolimits_\nu\langle u\star
f_\nu,g\rangle=\lim\nolimits_\nu\langle u,f_\nu \star
g\rangle=\langle u, f\star g\rangle=\langle u\star f, g\rangle
 \notag
 \end{equation}
and hence $h=u\star f$.  If one or other version of the closed
graph theorem~\cite{K,Sch} is applicable to  $E$, then these maps
are continuous and so belong to $\mathcal L(E)$.  This allows us
to define the   star products of the multipliers with  elements of
$E'$ by the formulas
\begin{equation}
\langle w\star u, f\rangle=(w,u\star f),\quad \langle v\star
w,f\rangle=\langle w,f\star v\rangle,
 \label{3.4}
\end{equation}
where $w\in E'$, $u\in {\mathcal M}_{\theta,L}(E)$, $v\in
{\mathcal M}_{\theta,R}(E)$, and $f$ ranges over $E$. So, the left
(right) Moyal multipliers of  $E$ serve as right (left)
multipliers for $E'$. Clearly, $E$ is contained in ${\mathcal
M}_{\theta,L}(E)$ as well as in ${\mathcal M}_{\theta,R}(E)$, and
operations~\eqref{3.4} extend operations~\eqref{3.1}. For fixed
$u$ and $v$, the maps $w\to w\star u$ and $w\to v\star w$ of $E'$
into itself are continuous because they are the transposes of the
maps $f\to v\star f$ and $f\to f\star u$ from  $E$ into $E$.

It is readily seen that  ${\mathcal M}_{\theta,L}(E)$ and
${\mathcal M}_{\theta,R}(E)$ are unital associative algebras with
respect to the star product. First we show that
$u_1,u_2\in{\mathcal M}_{\theta,L}(E)$ implies $u_1\star u_2\in
{\mathcal M}_{\theta,L}(E)$. Let $f,g\in E$. From~\eqref{3.1}, it
follows that
\begin{equation}
\langle(u_1\star u_2)\star f,g\rangle=\langle u_1\star u_2, f\star
g\rangle,
 \notag
\end{equation}
and by~\eqref{3.4} we have
\begin{equation}
\langle u_1\star u_2, f\star g\rangle=\langle u_1, u_2\star(f\star
g)\rangle.
 \notag
\end{equation}
Using~\eqref{3.3} and again~\eqref{3.1}, we obtain
\begin{equation}
\langle u_1, u_2\star(f\star g)\rangle=\langle u_1, (u_2\star
f)\star g\rangle=\langle u_1\star (u_2\star f), g\rangle,\quad
\text{for all $g\in E$}.
 \notag
\end{equation}
Therefore, $(u_1\star u_2)\star f=u_1\star (u_2\star f)\in E$ for
all $f\in E$ and hence $u_1\star u_2\in {\mathcal
M}_{\theta,L}(E)$. In a similar way, $v_1,v_2\in{\mathcal
M}_{\theta,R}(E)\Rightarrow v_1\star v_2\in {\mathcal M}_{\theta,
R}(E)$. Furthermore,
\begin{equation}
\langle(u_1\star u_2)\star u_3,f\rangle=\langle u_1\star u_2,
u_3\star f\rangle=\langle u_1, u_2\star(u_3\star f)\rangle=\langle
u_1, (u_2\star u_3)\star f\rangle=\langle u_1\star (u_2\star u_3),
f\rangle,
 \notag
\end{equation}
which proves the associativity of the algebra ${\mathcal
M}_{\theta,L}(E)$. We write $\mathds 1$ for the functional  $f\to
\int f(x)dx$. From~\eqref{3.*},
\begin{equation}
(\mathds 1\star f,g)=(f\star \mathds 1, g)=\int\!
f(x)g(x)dx,\qquad \text{for all $f,g\in E$}.
 \notag
\end{equation}
Hence $\mathds 1$ belongs to both  ${\mathcal M}_{\theta,L}(E)$
and ${\mathcal M}_{\theta,R}(E)$ and is the identity of these
algebras.

If  $E$ is invariant under the complex conjugation  $f\to f^*$ and
hence is an involutive algebra, then $E'$ also has an involution
$u\to u^*$, where $u^*$ is defined by
\begin{equation}
\langle u^*, f\rangle=\overline{\langle u,f^*\rangle}.
 \label{3.5}
\end{equation}
The involution~\eqref{3.5}  is an antilinear isomorphism of
${\mathcal M}_{\theta,L}(E)$ onto ${\mathcal M}_{\theta,R}(E)$.
Indeed, let $u$ be a left $\star$-multiplier of $E$ and let
$f,g\in E$. Then
\begin{equation}
\langle f\star u^*,g\rangle=\langle u^*,g\star
f\rangle=\overline{\langle u, (g\star
f)^*\rangle}=\overline{\langle u, f^*\star
g^*\rangle}=\overline{\langle u\star f^*, g^*\rangle}=(\langle
u\star f^*)^*, g\rangle,
 \notag
\end{equation}
and we see that the functional $f\star u^*$ is generated by the
test function $(u\star f^*)^*\in E$, hence $u^*\in {\mathcal
M}_{\theta,R}(E)$.

We let $\widehat E$ denote the Fourier transform of $E$ and equip
it with the topology  induced by the map $E\to\widehat E$. It
follows from definition~\eqref{1.2},  that
 \begin{equation}
\widehat{(f\star g)}(q)=(2\pi)^{-d}\int  \hat f(p)\hat
g(q-p)\,e^{\tfrac{i}{2}q\theta p}dp \label{3.6}
\end{equation}
for all $f,g\in S$. The integral expression on the right-hand side
of~\eqref{3.6} is called the twisted convolution product of  $\hat
f$ and $\hat g$. We denote\footnote{This notation follows
Kammerer~\cite{Kamm}, whereas in~\cite{Mail} the twisted
convolution operation was denoted by $ \ast_\theta$ and
in~\cite{G-BV1,G-BV2} by $\diamondsuit$.} this product by $\hat
f\mathbin{\hat{\star}_\theta}\hat g$ and, as before, omit the
explicit reference to $\theta$ whenever this cannot cause
confusion. Then~\eqref{3.6} takes the form
 \begin{equation}
\widehat{(f\star g)}=(2\pi)^{-d}\hat f\mathbin{\hat{\star}}\hat g,
\label{3.7}
\end{equation}
which is analogous to the familiar relation between the pointwise
multiplication and the ordinary convolution and turns into it at
$\theta=0$. If $E$ is a topological algebra under the Moyal
multiplication, then $\widehat E$ is a topological algebra under
the twisted convolution. From the foregoing it is clear that the
twisted convolution has a unique extension by continuity to the
case where one of factors is in $\widehat E'$. For $v\in\widehat
E'$ and $g\in\widehat E$, this extension is defined by
 \begin{equation}
v\hstar g=(2\pi)^d \mathcal F(\mathcal F^{-1}v\star \mathcal
F^{-1}g),\quad g\hstar v=(2\pi)^d\mathcal F(\mathcal F^{-1}g\star
\mathcal F^{-1}v). \label{3.8}
\end{equation}  The Fourier
transform is an isomorphism of the Moyal multiplier algebras
${\mathcal M}_{\theta,L}(E)$ and ${\mathcal M}_{\theta,R}(E)$
onto the twisted convolution multiplier algebras
\begin{align*}
{\mathcal C}_{\theta,L}(\widehat E)\defeq \{v\in\widehat E'\colon
v\hstar_\theta g\in\widehat E,\quad\text{for all  $g\in\widehat
E$}\}\\
\intertext{and}
 {\mathcal C}_{\theta,R}(\widehat E)\defeq \{v\in\widehat
 E'\colon  g\hstar_\theta
v\in\widehat E,\quad\text{for all $g\in\widehat E$}\}.
\end{align*}
Clearly, the Dirac $\delta$-function is the identity of the
algebras ${\mathcal C}_{\theta,L}(\widehat E)$ and ${\mathcal
C}_{\theta,R}(\widehat E)$.

\section{Moyal multipliers of the spaces of type S}

The spaces $S^\beta_\alpha$  are barrelled and fully complete.
Therefore, Ptak's version (see~\cite{Sch}, Sec.~IV.8.5) of the
closed graph theorem is applicable to their linear endomorphisms.
Each of them is dense in $S$ and, by Theorem~1 of~\cite{I}, the
spaces $S^\beta_\alpha$ with $\alpha\ge\beta$ are topological
algebras under the Moyal multiplication. Therefore,  it follows
from the above general consideration that the algebras ${\mathcal
M}_{\theta,L}(S^\beta_\alpha)$, ${\mathcal
M}_{\theta,R}(S^\beta_\alpha)$, ${\mathcal
C}_{\theta,L}(S_\beta^\alpha)$, and ${\mathcal
C}_{\theta,R}(S_\beta^\alpha)$ are well defined for
$\alpha\ge\beta$. To describe their properties  we need the
following lemma.

 {\bf Lemma~1:} {\it Let $\alpha\ge \beta$ and let $e$
be a function in $S^\beta_\alpha(\oR^d)$ such that $e(0)=1$. Let
$e_\nu(x)=e(x/\nu)$, $\nu=1,2,\dots$. Then either of the operator
sequences $f\to e_\nu\star f$ and $f\to f\star e_\nu$ converges to
the identity map of $S^\beta_\alpha$ uniformly on the bounded
subsets of $S^\beta_\alpha$.

Proof.} From~\eqref{1.2}, we have
\begin{equation}
(f\star e_\nu)(x)=\int  \omega_\nu(q) f(x-\tfrac12\theta
q)\,e^{iqx} dq,
 \label{4.1}
\end{equation}
where $\omega_\nu(q)\defeq(\nu/2\pi)^{d}\hat e(\nu q)$. The
function $\hat e$ belongs to $S_\beta^\alpha$, hence $\omega_\nu$
satisfies the inequality
\begin{equation}
|\omega_\nu(q)|\le C_0 \nu^d e^{-|\nu q/B_0|^{1/\beta}}
  \label{4.2}
\end{equation}
with some positive constants $C_0$,$ B_0$. Since $\int\omega_\nu
(q)dq=e(0)=1$, it follows from~\eqref{4.1}  that
\begin{multline}
\partial^\on(f\star e_\nu-f)(x)=\int\omega_\nu(q)\left(e^{iqx}\partial^\on
f(x-\tfrac12\theta q)-\partial^\on f(x)\right)dq\, +\\
+\int\omega_\nu(q)\sum_{\om\ne\mathsf 0}\binom{\on}\om (iq)^\om
e^{iqx}\partial^{\on-\om} f(x-\tfrac12\theta q)\,dq. \label{4.3}
\end{multline}
We define $F_x(q)= e^{iqx}\partial^\on f(x-\tfrac12\theta q)$. By
the mean value theorem,
\begin{multline}
|F_x(q)-F_x(0)|\le |q|\sup_{|p|\le
|q|}\sum_{j=1}^d\left|\frac{\partial F_x(p)}{\partial p_j}\right|\le\\
\le|q|\sup_{|p|\le |q|}\left\{d|x||\partial^\on f(x-\tfrac12\theta
p)|+ \frac12\sum_{i,j}|\theta^{ij}
\partial_{x^i}\partial^\on f(x-\tfrac12\theta p)|\right\}.
\label{4.4}
\end{multline}
Using definition~\eqref{2.2} and inequality~\eqref{2.8}, we obtain
\begin{multline}
|x|\,|\partial^\on f(x-\tfrac12\theta p)|\le
\|f\|_{A,B}B^{|\on|}\on^{\beta\on}|x| e^{-|(x-\theta
p/2)/A|^{1/\alpha}}\le\\ \le C_{A'}\|f\|_{A,B}
B^{|\on|}\on^{\beta\on}e^{-|x/A'|^{1/\alpha}+(|\theta||
q/2A|)^{1/\alpha}}, \notag
\end{multline}
where $|\theta|=\sum|\theta^{ij}|$ and $A'$ is an arbitrary
constant greater than $2A$. The sum in the braces on the
right-hand side of~\eqref{4.4} is estimated in a similar manner,
using the inequality $(n+1)^{\beta (n+1)}\le C_\epsilon
(1+\epsilon)^nn^{\beta n}$,  and this yields
\begin{equation}
|F_x(q)-F_x(0)|\le C\|f\|_{A,B}
B'^{|\on|}\on^{\beta\on}|q|\,e^{-|x/A'|^{1/\alpha}+(|\theta||
q/2A|)^{1/\alpha}},
 \label{4.5}
\end{equation}
where  $B'>B$ and can be taken arbitrarily close to $B$. We let
$I_\nu^{(1)}(x)$ and $I_\nu^{(2)}(x)$ denote the integrals on the
right-hand side of~\eqref{4.3}. Since $\alpha\ge \beta$,
from~\eqref{4.2} and \eqref{4.5} it follows that for
$\nu>B_0|\theta|/A$, the first integral satisfies the estimate
\begin{equation}
 |I_\nu^{(1)}(x)|\le\frac{1}{\nu}C'\|f\|_{A,B}
B'^{|\on|}\on^{\beta\on}e^{-|x/A'|^{1/\alpha}},
 \label{4.6}
\end{equation}
where  $C'=C C_0\int |q|
e^{-|q/B_0|^{1/\beta}+|q/2B_0|^{1/\beta}}dq$. To estimate the
second integral we take into account that if $\om\ne \mathsf 0$,
then
\begin{equation}
 |q^\om|\le |q|\,B_1^{|\om|-1}
\om^{\beta\om}\prod_{j=1}^de^{(\beta/e)|q_j/B_1|^{1/\beta}}\quad\text{for
each $B_1>0$}.
 \label{4.*}
\end{equation}
This is obtained by writing $q^\om=|q|q^{\om'}$, where
$|\om'|=|\om|-1$, and evaluating $\sup_\om|q^\om|/\om^{\beta\om}$.
Using~\eqref{4.*} together with~\eqref{2.8} and \eqref{4.2}, we
find that
\begin{multline}
 |I_\nu^{(2)}(x)|\le \frac{C_0}{B_1}\|f\|_{A,B}e^{-|x/2A|^{1/\alpha}}
 \sum_{\om}\binom{\on}\om
B_1^\om B^{|\on-\om|}\om^{\beta\om}(\on-\om)^{\beta(\on-\om)}\\
\times\int |q|e^{-|\nu
q/B_0|^{1/\beta}+(d\beta/e)|q/B_1|^{1/\beta}+(|\theta||
q/2A|)^{1/\beta}}\nu^d dq
 \notag
\end{multline}
Hence, if $\nu$ is so large that
$\nu^{1/\beta}>(d\beta/e)(2B_0/B_1)^{1/\beta}+(|\theta|B_0/A)^{1/\beta}$,
we have
\begin{equation}
 |I_\nu^{(2)}(x)|\le\frac{1}{\nu}\frac{C^\prime}{B_1}\|f\|_{A,B}
(B+B_1)^{|\on|}\on^{\beta\on}e^{-|x/2A|^{1/\alpha}}.
 \label{4.7}
\end{equation}
Now we recall that the inductive limit $\injlim_{A,B\to\infty}
S^{\beta,B}_{\alpha,A}$ is regular. In other words for any bounded
subset $Q$ of $S^\beta_\alpha$, there are $A$ and $B$ such that
$Q$ is contained in $S^{\beta,B}_{\alpha,A}$ and is bounded in its
norm. Let $A'>2A$ and $B'>B$ as before.  Using~\eqref{4.6}
and~\eqref{4.7} with $B_1=B'-B$, we conclude that if $f$ ranges
over $Q$ and $\nu$ is large enough, then the functions $f\star
e_\nu-f$ belong to $S^{\beta,B'}_{\alpha,A'}$  and $\sup_{f\in
Q}\|f\star e_\nu-f\|_{A',B'}\to 0$ as $\nu\to 0$. Thus, the
sequence of the operators of right $\star$-multiplication by
$e_\nu$ converges to the unit operator uniformly on every bounded
subset of $S^\beta_\alpha$. The argument for the case of left
$\star$-multiplication is analogous.

{\bf Theorem~3:} {\it The algebra ${\mathcal
M}_{\theta,L}(S^\beta_\alpha)$, where $\alpha\ge\beta$,  can be
canonically identified with the closure in $\mathcal
L(S^\beta_\alpha)$ of the set of operators of left
$\star_\theta$-multiplication by elements of $S^\beta_\alpha$. An
analogous statement is true for ${\mathcal
M}_{\theta,R}(S^\beta_\alpha)$.

Proof.} Let  $u\in {\mathcal M}_{\theta,L}(S^\beta_\alpha)$ and
let $U\in \mathcal L(S^\beta_\alpha)$ be the operator  that takes
each function $f\in S^\beta_\alpha$
 to $u\star f$, so that
\begin{equation}
\langle u\star f,g\rangle=\int U(f)g \,dx,\qquad \text{ for all
$g\in S^\beta_\alpha$}.
 \label{4.8}
 \end{equation}
If $u_1$ and $u_2$ determine the same operator, then $\langle u_1,
f\star g\rangle=\langle u_2, f\star g\rangle$ for all $f,g\in
S^\beta_\alpha$.  Lemma~1 shows in particular that the set
 $\{f\star g\colon f,g\in
S^\beta_\alpha\}$ is dense in $S^\beta_\alpha$. Therefore the map
\begin{equation}
{\mathcal M}_{\theta,L}(S^\beta_\alpha)\to\mathcal
L(S^\beta_\alpha)\colon u\to U
 \label{4.9}
\end{equation}
is one-to-one.  Every operator belonging to its image has the
property
\begin{equation}
U(h\star f)=U(h)\star f,
 \label{4.10}
 \end{equation}
where $f$ and $h$ are arbitrary elements of $S^\beta_\alpha$.
Indeed, from~\eqref{3.2} and~\eqref{3.3} it follows that
\begin{equation}
\int U(h\star f) g \,dx=\langle u\star(h\star f),g\rangle=\langle
u\star h, f\star g\rangle=\int U(h)(f\star g) \,dx=\int (U(h)\star
f) g \,dx,
 \notag
 \end{equation}
for all $g\in S^\beta_\alpha$.  Let $e_\nu$ be the sequence
defined in Lemma~1. From~\eqref{4.10}, we see that the operator
$U_\nu\colon f\to U(e_\nu\star f)$ consists in the left
$\star$-multiplication by $U(e_\nu)$ and  by Lemma~1, the operator
sequence $U_\nu$ converges to $U$ in $\mathcal L(S^\beta_\alpha)$
as $\nu\to\infty$.

On the other hand, to each $T\in \mathcal L(S^\beta_\alpha)$ we
can assign a functional $t\in (S^\beta_\alpha)'$ by setting
\begin{equation}
\langle t,f\rangle=\int T(f)\,dx,\qquad f\in S^\beta_\alpha.
 \label{4.11}
 \end{equation}
The map $T\to t$ from $\mathcal L(S^\beta_\alpha)$ into
$(S^\beta_\alpha)'$ is continuous by the definition of topologies
of these spaces. Using~\eqref{4.10} again, we see that the
composition of~\eqref{4.9} and this map is the identity map of
${\mathcal M}_{\theta,L}(S^\beta_\alpha)$. Let $T=\lim_\nu T_\nu$,
where $T_\nu$ is the operator of left $\star$-multiplication by
$g_\nu\in S^\beta_\alpha$. Then  we have
\begin{equation}
T(h\star f)=\lim_\nu g_\nu\star(h\star f)=\lim_\nu (g_\nu\star
h)\star f= T(h)\star f,\qquad \text{for all $h\in
S^\beta_\alpha$}.
 \notag
 \end{equation}
This implies that the functional $t$ corresponding to $T$
by~\eqref{4.11} satisfies  $t\star h=T(h)$ and hence belongs to
${\mathcal M}_{\theta,L}(S^\beta_\alpha)$. The algebra ${\mathcal
M}_{\theta,R}(S^\beta_\alpha)$ can be considered in a similar way,
and this completes the proof.

Theorem~3 shows that the natural topology on the left and right
multiplier algebras of $S^\beta_\alpha$ is the topology induced by
that of $\mathcal L(S^\beta_\alpha)$. Now we define analogs of the
involutive algebra of $\star$-multipliers introduced by
Antonets~\cite{A1,A2} for $S$. Namely, we consider the
intersection
\begin{equation}
{\mathcal M}_\theta(S^\beta_\alpha)={\mathcal M}_{\theta,
L}(S^\beta_\alpha)\cap{\mathcal M}_{\theta, R}(S^\beta_\alpha).
\label{4.12}
 \end{equation}
The Moyal product of two its elements $u$ and $v$ can be defined
by
\begin{equation}
\langle u\star v,f\rangle=\langle u,v\star f\rangle,
 \label{4.13}
 \end{equation}
or,  with the same right, by
\begin{equation}
\langle u\star v,f\rangle=\langle v,f\star u \rangle. \label{4.14}
 \end{equation}
To prove that these definitions are equivalent, we again use the
sequence $e_\nu$ of Lemma~1. By~\eqref{3.1} and~\eqref{3.3}, we
have
\begin{gather}
\langle e_\nu \star u,v\star f\rangle= \langle u,(v\star f)\star
e_\nu \rangle\to\langle u,v\star f\rangle\quad
(\nu\to\infty),\notag
\\
\langle v, f\star(e_\nu \star u)\rangle= \langle v, (f\star e_\nu)
\star u\rangle\to\langle v,f\star u \rangle \quad (\nu\to\infty).
\notag
\end{gather}
It remains to note that $\langle g,v\star f\rangle=\langle v,
f\star g\rangle$ for any $g\in S^\beta_\alpha$ and, in particular,
for $g=e_\nu \star u$. This is obtained by passing to the limit as
$\nu\to\infty$ in the equality  $\langle g,(v\star e_\nu)\star
f\rangle=\langle v\star e_\nu, f\star g\rangle$ which holds
by~\eqref{3.2}. The space ${\mathcal M}_\theta(S^\beta_\alpha)$
can be made into a locally convex space by giving it the least
upper bound of the topologies induced by those of  ${\mathcal
M}_{\theta,L}(S^\beta_\alpha)$ and ${\mathcal
M}_{\theta,R}(S^\beta_\alpha)$.

 Before
formulating the next theorem we recall that in the theory of
bilinear maps of locally convex spaces, a large role is played by
the notion of  $({\mathfrak B}_1, {\mathfrak
B}_2)$-hypocontinuity~\cite{Sch} which takes an intermediate
position between continuity and separate continuity. In the case
where ${\mathfrak B}_1$ and ${\mathfrak B}_2$ are families of all
bounded subsets of the spaces $E_1$ and $E_2$ on whose direct
product a bilinear map is defined, this property is often termed
hypocontinuity for short.

{\bf Theorem~4:} {\it Under the condition $\alpha\ge\beta$,
${\mathcal M}_\theta(S^\beta_\alpha)$  is a complete nuclear
semireflexive unital *-algebra with hypocontinuous multiplication
and continuous involution. The space $(S^\beta_\alpha)'$ is an
${\mathcal M}_\theta(S^\beta_\alpha)$-bimodule with hypocontinuous
operations $(v,w)\to w\star v$ and $(v,w)\to v\star w$, where
$w\in (S^\beta_\alpha)'$, and $v\in {\mathcal
M}_\theta(S^\beta_\alpha)$.

Proof.}  We noticed already that  $\mathcal L(S^\beta_\alpha)$ has
the topological properties listed above. Its closed subspaces
${\mathcal M}_{\theta,L}(S^\beta_\alpha)$ and ${\mathcal
M}_{\theta,R}(S^\beta_\alpha)$ as well as their intersection also
are  complete, nuclear, and semireflexive by the well known
hereditary properties~\cite{Sch}.

From the definition of the topology on ${\mathcal
M}_\theta(S^\beta_\alpha)$, it immediately follows that the
multiplication in this algebra is separately continuous. Indeed, a
base of neighborhoods of 0 in   ${\mathcal
M}_\theta(S^\beta_\alpha)$ is formed by the sets of the form
$\mathcal V_{Q,\mathcal U}=\{v\colon (v\star Q)\cup(Q\star
v)\subset\mathcal U\}$, with $Q$  a bounded subset of
$S^\beta_\alpha$ and $\mathcal U$  a neighborhood of 0 in
$S^\beta_\alpha$. For any fixed $u\in {\mathcal
M}_\theta(S^\beta_\alpha)$ and for each neighborhood $\mathcal U$,
there is a  neighborhood $\mathcal U_1$ such that $u\star \mathcal
U_1\subset\mathcal  U$. Taking $Q'=Q\cup (Q\star u)$ and $\mathcal
U'=\mathcal U\cap\mathcal U_1$, we have the implication
 \begin{equation} v\in \mathcal V_{Q',\mathcal U'}
\quad\Longrightarrow\quad u\star v \in \mathcal V_{Q,\mathcal
U}\,, \label{implication}
 \end{equation}
which shows that the map $v\to u\star v$ of ${\mathcal
M}_\theta(S^\beta_\alpha)$ into itself is continuous. Similarly,
the map $u\to u\star v$ is continuous for every fixed $v$. Now let
$u$ range over a bounded subset $\mathcal Q$ of ${\mathcal
M}_\theta(S^\beta_\alpha)$. Then the set $Q\star\mathcal Q$ is
bounded in $S^\beta_\alpha$. Since the space $S^\beta_\alpha$ is
barrelled,   we may apply the general principle of uniform
convergence~\cite{Sch} and conclude that there is a neighborhood
$\mathcal U_2$ such that $\mathcal Q\star\mathcal
U_2\subset\mathcal U$. Taking this time $Q'=Q\cup (Q\star\mathcal
Q)$ and $\mathcal U'=\mathcal U\cap\mathcal U_2$, we again have
implication~\eqref{implication}, hence the bilinear map $(u,v)\to
u\star v$ is $\mathfrak B_1$-hypocontinuous. Analogously, it is
$\mathfrak B_2$-hypocontinuous.

A base of neighborhoods for $S^\beta_\alpha$ can obviously  be
formed of sets invariant under the involution $f\to f^*$, and
every bounded subset of $S^\beta_\alpha$ is contained in an
invariant bounded subset. Therefore, the family of sets of the
form  $\{u\colon f\star u\in\mathcal U,\,u\star f\in\mathcal
U,\,\forall f\in Q\}$, with   $\mathcal U$  an invariant
neighborhood and $Q$  an invariant bounded set in
$S^\beta_\alpha$, forms a base of neighborhoods for ${\mathcal
M}_\theta(S^\beta_\alpha)$ which is invariant under the map $u\to
u^*$.

To prove that $(S^\beta_\alpha)'$ is a bimodule over the algebra
${\mathcal M}_\theta(S^\beta_\alpha)$, it suffices to show that
\begin{equation}
(u\star v)\star w=u\star (v\star w),\quad (u\star w)\star v=u\star
(w\star v),\quad (w\star u)\star v=w\star (u\star v),
 \label{4.15}
 \end{equation}
for all $w\in (S^\beta_\alpha)'$ and for any $u,v\in {\mathcal
M}_\theta(S^\beta_\alpha)$. These associativity relations follow
immediately from analogous relations for the action of the algebra
${\mathcal M}_\theta(S^\beta_\alpha)$ on  test functions. In
particular, for any $f\in S^\beta_\alpha$, we have the chain of
equalities
\begin{equation}
\langle(u\star v)\star w,f\rangle=\langle w,f\star(u\star
v)\rangle= \langle w,(f\star u)\star v\rangle=\langle v\star
w,f\star u\rangle=\langle u\star (v\star w), f\rangle,
 \notag
 \end{equation}
which proves the first of relations~\eqref{4.15}.  As was already
noted in Sec.~III, the maps $w\to w\star v$ and $w\to v\star w$,
being transposes of continuous maps, are  continuous. Now we fix
$w\in (S^\beta_\alpha)'$ and show that the maps $v\to v\star w$
and $v\to v\star w$ of ${\mathcal M}_\theta(S^\beta_\alpha)$ into
$(S^\beta_\alpha)'$ are also continuous. By the definition of the
strong topology, every neighborhood of 0 in $(S^\beta_\alpha)'$
contains a set of the form $Q^\circ=\{t\colon \sup_{f\in
Q}|\langle t,f\rangle|\le 1\}$, where $Q$  is a bounded subset of
$S^\beta_\alpha$. Because the functional $w$ is continuous, there
is a neighborhood  $\mathcal U$ in $S^\beta_\alpha$ such that
$\sup_{f\in\mathcal  U}|\langle w,f\rangle|\le 1$.   Clearly,
$v\in \mathcal V_{Q,\mathcal U}$ implies $w\star v \in Q^\circ$
and $v\star w\in Q^\circ$. Hence the bilinear maps $(v,w)\to
w\star v$ and $(v,w)\to v\star w$ are separately continuous.
Moreover, since the space $(S^\beta_\alpha)'$ is barrelled, they
are $\mathfrak B_1$-hypocontinuous. Now let $w$ range over a
bounded set $B\subset (S^\beta_\alpha)'$. Because $S^\beta_\alpha$
is barrelled, this set of functionals is equicontinuous, i.e.,
there is a neighborhood $\mathcal U$ of 0 in $S^\beta_\alpha$ such
that $\sup_{f\in\mathcal  U}|\langle w,f\rangle|\le 1$ for all
$w\in B$. Therefore, the bilinear maps under consideration are
$\mathfrak B_2$-hypocontinuous, and this completes the proof of
Theorem~4.

 In concluding this section, we note that the
Fourier transform is an isomorphism of  the Moyal multiplier
algebra ${\mathcal M}_\theta(S^\beta_\alpha)$, where
$\alpha\ge\beta$,  onto the algebra
\begin{equation}
{\mathcal C}_\theta(S_\beta^\alpha)={\mathcal C}_{\theta,
L}(S_\beta^\alpha)\cap{\mathcal C}_{\theta, R}(S_\beta^\alpha)
\label{4.16}
 \end{equation}
and $(S_\beta^\alpha)'$ is  a unital ${\mathcal
C}_\theta(S_\beta^\alpha)$-bimodule with hypocontinuous
operations.

\section{Smoothness and growth properties of the twisted\\
convolution product}

We now focus our attention on the smoothness and growth properties
of the twisted convolution products $g\hstar v$ and $v\hstar g$,
where  $g\in S^\alpha_\beta$, $v\in (S^\alpha_\beta)'$, and
$\alpha\ge\beta$. We will show that these properties are not too
much different from those of the undeformed convolution product
$g*v$. For this purpose, we use the following lemma.

{\bf Lemma~2:} {\it If $f\in S_{\alpha, A}^{\beta, B}(\oR^d)$,
then its Fourier transform $\hat f$ belongs to $S^{\alpha,
rA}_{\beta, rB}(\oR^d)$, where the number $r$ depends only on
$\alpha$, $\beta$, and $d$. The  map  $S_{\alpha, A}^{\beta, B}\to
S^{\alpha, rA}_{\beta, rB}\colon f\to\hat f$ is bounded, i.e.,
there is a  constant $C$ such that }
\begin{equation}
\|\hat f\|_{rB, rA}\le C\|f\|_{A,B}, \quad \text {for all $f\in
S_{\alpha, A}^{\beta, B}$}.
 \label{5.1}
\end{equation}

The proof of Lemma~2 is presented in Appendix. It is based on  an
inequality proposed in~\cite{S82}, which allows us to simplify
considerably the theory~\cite{GS2} of Fourier transform on the
spaces of type $S$.

{\bf Theorem~5:} {\it Suppose that $\alpha\ge \beta$, $g\in
S^\alpha_\beta$ and $v\in (S^\alpha_\beta)'$. Then the twisted
convolution products $v\hstar g$ and $g\hstar v$ belong to the
space $M(S^\alpha_\beta)$ and can be written as
\begin{equation}
(v\hstar g)(q)= \left\langle
v,g(k-\cdot)e^{\tfrac{i}{2}q\theta(\cdot)}\right\rangle,\quad
(g\hstar v)(q)= \left\langle
v,g(q-\cdot)e^{-\tfrac{i}{2}q\theta(\cdot)}\right\rangle.
 \label{5.2}
\end{equation}
The maps $(g,v)\to v\hstar g$ and $(g,v)\to g\hstar v$ from
$S^\alpha_\beta\times (S^\alpha_\beta)'$ into $M(S^\alpha_\beta)$
are hypocontinuous.

Proof.} If  $v$ is a regular functional generated by a function in
$S$, then~\eqref{5.2} is obviously consistent with the above
definition of the twisted convolution product for elements of the
Schwartz space. The functions on the right-hand sides
of~\eqref{5.2} are well defined because  $S^\alpha_\beta$ is
invariant under the  reflection and  translations of $R^d$, and
under the multiplication by $e^{\pm\tfrac{i}{2}q\theta (\cdot)}$
which is equivalent to a translation of the Fourier transforms in
$S_\alpha^\beta$. It suffices to show that these functions belong
to $M(S^\alpha_\beta)$ and depend continuously on $v$. Then we may
state that the products defined by~\eqref{5.2} extend continuously
the twisted convolution multiplication to the case where one of
factors belongs to $(S^\alpha_\beta)'$, because
$M(S^\alpha_\beta)$ is contained in $(S^\alpha_\beta)'$ and its
topology is stronger than that induced from $(S^\alpha_\beta)'$.
Such an extension is unique because $S^\alpha_\beta$ is dense in
$(S^\alpha_\beta)'$, hence  functions~\eqref{5.2} coincide with
the products defined by~\eqref{3.8}.

Now we introduce the notation
\begin{equation}
g^{\pm}_q(p)= g(q-p)e^{\pm\tfrac{i}{2}q\theta p}.
 \label{5.3}
\end{equation}
The functions $\langle v,g^{\pm}_q\rangle$ are infinitely
differentiable and
\begin{equation}
\partial_q^\on (v,g^{\pm}_q)=(v,\partial_q^\on g^{\pm}_q)
 \label{5.4}
\end{equation}
for any $d$-tuple  $\on\in \oZ^d_+$. To prove formula~\eqref{5.4},
we first observe that  the operators $\partial^\on$ are defined
and continuous in $S^\alpha_\beta$ because by
definition~\eqref{2.2} and the inequality $(n+m)^{(n+m)}\le
2^{(n+m)}n^{n}m^{m}$ we have
\begin{equation}
\|\partial^\on g\|_{B,2^\alpha A}\le(2^\alpha
A)^{|\on|}\on^{\alpha \on}\|g||_{B,A},
 \label{5.5}
\end{equation}
for all $g\in S^{\alpha,A}_{\beta,B}$.  The translation operators
$T_q\colon g(p)\to g(p-q)$   are also defined and continuous in
$S^\alpha_\beta$. Using the inequality
$-|p-q|^{1/\beta}\le-|p/2|^{1/\beta}+|q|^{1/\beta}$, we obtain
\begin{equation}
\|T_q g\|_{2B,A}\le e^{|q/B|^{1/\beta}}\|g\|_{B,A}
 \label{5.7}
\end{equation}
and conclude that the set of functions  $T_q g$, where $|q|\le
c<\infty$, is bounded in $S^\alpha_\beta$ for any fixed $g$. Since
$S^\alpha_\beta$ is a Montel space, it follows that    $T_q$ is
continuous in $q$. As shown in Sec.~III.3.3 of~\cite{GS2}, these
properties of differentiation and translation in
$S^{\alpha,A}_{\beta,B}$ imply that the difference quotient
$(T_{-q_j}f-f)/q_j$, where $T_{q_j}$ is the operator of
translation in the $j$th coordinate, converges to $\partial_jf$ in
$S^{\alpha,A}_{\beta,B}$. The operator of multiplication by
$e^{\pm\tfrac{i}{2}q\theta p}$  is  also  strongly differentiable
in $q$, because the (inverse) Fourier transform converts it into
the translation by  $\pm \tfrac{1}{2}\theta q$ in
$S_\alpha^\beta$.  As a result, we arrive at~\eqref{5.4}.

From~\eqref{5.4}, it follows that
\begin{equation}
|\partial^\on(v\hstar g)(q)|\le \|v\|_{B,A}\|\partial^\on_q
g^+_q\|_{B,A}\quad \text{and}\quad |\partial^\on(g\hstar v)(q)|\le
\|v\|_{B,A}\|\partial^\on_q g^-_q\|_{B,A},
 \label{5.8}
\end{equation}
where by the Leibniz rule,
\begin{equation} \|\partial^\on_q
g^\pm_q\|_{B,A}\le\sum_\om\binom{\on}{\om}\frac{1}{2^{|\om|}}
\left\|(\theta p)^\om e^{\pm\tfrac{i}{2}q\theta
p}\partial^{\on-\om}g(q-p)\right\|_{B,A}.
 \label{5.9}
\end{equation}
Let $g\in S^{\alpha,A_0}_{\beta,B_0}$ and let $A_1=2^\alpha A_0$,
$B_1=2B_0$. Using~\eqref{5.5} and~\eqref{5.7}, we obtain
\begin{equation}
\|\partial^{\on-\om}g(q-p)\|_{B_1,A_1}\le \|g\|_{B_0,A_0}
(2^\alpha
A_0)^{|\on-\om|}(\on-\om)^{\alpha(\on-\om)}e^{|q/B_0|^{1/\beta}}.
 \label{5.10}
\end{equation}
For any $h\in S^\alpha_\beta$, we have the inequality
\begin{equation}
\|(\theta p)^\om e^{\pm\tfrac{i}{2}q\theta p}h(p)\|_{B,A}\le
|\theta|^{|\om|}\max_{|\ok|=|\om|}\|p^\ok
e^{\pm\tfrac{i}{2}q\theta p}h(p)\|_{B,A},
 \label{5.11}
\end{equation}
where the norms are finite if $A$  and $B$ are large enough.
Furthermore,
\begin{equation}
\mathcal F^{-1}\left(p^\ok e^{\pm\tfrac{i}{2}q\theta
p}h(p)\right)=\frac{1}{(2\pi)^d}\int p^\ok
e^{ip(x\mp\tfrac{1}{2}\theta q)}h(p)dp=(-i)^{|\ok|}\partial^\ok
f(x\mp\tfrac12\theta q),
 \notag
\end{equation}
where $\hat f =h$. If $h\in S^{\alpha,A_1}_{\beta,B_1}$, then
$\|f\|_{rA_1,rB_1}\le C\|h\|_{B_1,A_1}$ by Lemma~2. (More exactly,
by an analog of this lemma for $\mathcal F^{-1}$.) Next we use
analogs of inequalities~\eqref{5.5} and~\eqref{5.7} for functions
in $S_\alpha^\beta$. This gives the estimate
\begin{multline}
\|\partial^\ok f(x\mp\tfrac12\theta q)\|_{2rA_1,2^\beta rB_1} \le
\|f(x\mp\tfrac12\theta q)\|_{2rA_1, rB_1} (2^\beta
rB_1)^{|\ok|}\ok^{\beta\ok}\\
\le C \|h\|_{B_1,A_1} (2^\beta
rB_1)^{|\ok|}\ok^{\beta\ok}e^{|\theta q/(2rA_1)|^{1/\alpha}}.
 \label{5.12}
\end{multline}
By Lemma~2, we have
\begin{equation}
\|p^\ok e^{\pm\tfrac{i}{2}q\theta p}h(p)\|_{B,A}\le
C'\|\partial^\ok f(x\mp\tfrac12\theta q)\|_{2rA_1,2^\beta
rB_1},\quad \text{where $A=2rr'A_1$, $B=2^\beta rr'B_1$}.
\label{5.13}
 \end{equation}
Combining~\eqref{5.11} and \eqref{5.12} with~\eqref{5.13} and
taking into account that
$\max_{|\ok|=|\om|}\ok^\ok=|\om|^{|\om|}\le d^{|\om|}\om^{\om}$,
we get
\begin{equation}
\|(\theta p)^\om e^{\pm\tfrac{i}{2}q\theta p}h(p)\|_{B,A}\le
C^{\prime\prime}\|h\|_{B_1,A_1} (2^\beta d^\beta|\theta|
rB_1)^{|\om|}\om^{\beta\om}e^{|\theta q/(2rA_1)|^{1/\alpha}}.
 \notag
 \end{equation}
Substituting here $h(p)=\partial^{\on-\om}g(q-p)$  and
using~\eqref{5.9} and~\eqref{5.10},  we conclude that
\begin{multline}
\|\partial^\on_q g^\pm_q\|_{B,A}\le
C^{\prime\prime}\|g\|_{B_0,A_0}\sum_\om\binom{\on}{\om}(2^\beta
d^\beta|\theta| rB_0)^{|\om|} (2^\alpha
A_0)^{|\on-\om|}\om^{\beta\om}(\on-\om)^{\alpha(\on-\om)}\\
\times e^{|q/B_0|^{1/\beta}+\tfrac{1}{2}|\theta
q/(2rA_0)|^{1/\alpha}},
 \notag
 \end{multline}
where $A=2^{\alpha+1}rr'A_0$ and $B=2^{\beta+1} rr'B_0$. If
$\alpha\ge\beta$, then
$\om^{\beta\om}(\on-\om)^{\alpha(\on-\om)}\le \on^{\alpha\on}$,
and we arrive at the inequality
\begin{equation}
\|\partial^\on_q g^\pm_q\|_{B,A}\le
C^{\prime\prime}\|g\|_{B_0,A_0}\left(2^\beta d^\beta|\theta|
rB_0+2^\alpha A_0\right)^{|\on|}\on^{\alpha\on}
e^{|q/B_0|^{1/\beta}+\tfrac{1}{2}|\theta q/(2rA_0)|^{1/\beta}}.
 \label{5.14}
 \end{equation}
A similar inequality obviously holds for any $A'_0>A_0$,
$B'_0>B_0$ and their corresponding $A'=2^{\alpha+1}rr'A'_0$,
$B'=2^{\beta+1} rr'B'_0$, because
$S^{\alpha,A_0}_{\beta,B_0}\subset S^{\alpha,A'_0}_{\beta,B'_0}$.
Let $A'_0=A_0+|\theta|R/2r$ and $B'_0=B_0+2^\beta R$, where $R$ is
a positive number. Then
\begin{equation}
|q/B'_0|^{1/\beta}+\tfrac{1}{2}|\theta q/(2rA'_0)|^{1/\beta}\le
|q/R|^{1/\beta},
 \notag
 \end{equation}
and we obtain
\begin{equation}
\|\partial^\on_q g^\pm_q\|_{B',A'}\le
C^{\prime\prime\prime}\|g\|_{B_0,A_0}A_R^{|\on|}\on^{\alpha\on}
e^{|q/R|^{1/\beta}}.
 \label{5.15}
 \end{equation}
where $A_R= 2^\alpha A_0 +s|\theta|(B_0+R)$ with a positive
coefficient  $s$ depending only on $\alpha$, $\beta$ and $d$,
whose explicit form is irrelevant. From~\eqref{5.8}
and~\eqref{5.15}, it follows that
\begin{equation}
\|v\hstar g \|_{-R, A_R}\equiv\sup_{q,\on}\left|\,
\frac{\partial^\on (v\hstar g )
(q)}{A_R^{|\on|}\on^{\alpha\on}}\,e^{-|q/R|^{1/\beta}}\right| \le
C^{\prime\prime\prime}\|v\|_{B',A'}\|g\|_{B_0,A_0}.
 \label{5.16}
\end{equation}
An analogous inequality holds for $g\hstar v$. We conclude that
the functions  $v\hstar g$ and $g\hstar v$ belong to any space
$E^{\alpha, A_R}_{\beta, R}$ with $R>0$,  and {\it a fortiori}  to
$M(S_\beta^\alpha)$. From the form of the right-hand side
of~\eqref{5.16}, it is clear that the maps  $(g,v)\to g\hstar v$
and $(g,v)\to v\hstar g$ from $S^\alpha_\beta\times
(S^\alpha_\beta)'$ into $M(S^\alpha_\beta)$ are separately
continuous. Since the spaces $S^\alpha_\beta$ and
$(S^\alpha_\beta)'$ are barrelled, this amounts to saying that
these maps are hypocontinuous, and thus we have proved Theorem~5.

{\it Corollary: If $\alpha\ge \beta$, $f\in S_\alpha^\beta$ and
$u\in (S_\alpha^\beta)'$, then the functionals $u\star f$ and
$f\star u$ belong to $C(S^\alpha_\beta)$. The maps $(f,u)\to
u\star f$ and $(f,u)\to f\star u$ from $S_\alpha^\beta\times
(S_\alpha^\beta)'$ into $C(S_\alpha^\beta)$ are hypocontinuous.}

We notice that for $\theta=0$, the constant $A_R$ equals $2^\alpha
A_0$ and becomes independent of $R$. In this case,
formula~\eqref{5.16} shows that for all $g\in S^\alpha_\beta$ and
$v\in (S^\alpha_\beta)'$, the usual convolution $v\ast g$  belongs
to the space $C'(S^\alpha_\beta)$ defined by~\eqref{2.7}, which is
smaller than $M(S^\alpha_\beta)$. The  deformed convolution need
not belong to  $C'(S^\alpha_\beta)$  but is contained in
$M(S^\alpha_\beta)$ for any $\theta$. This  is analogous to the
result obtained previously~\cite{G-BV1} in the framework of
tempered distributions, where the roles of $C'(S^\alpha_\beta)$
and $M(S^\alpha_\beta)$ are played respectively by the spaces
$\mathcal O_C=C'(S)$ and $\mathcal O_M=M(S)$. The analogy is even
more complete. The linear dependence of $A_R$ on $R$, established
by Theorem~5 for $\theta\ne 0$, is an analog of the fact that for
$g\in S$ and $v\in S'$, the twisted convolutions $v\hstar g$ and
$g\hstar v$ belong to the space denoted in~\cite{G-BV1} by
$\mathcal O_T$. This space is smaller than the Schwartz multiplier
space $\mathcal O_M$ and consists of all smooth functions with
polynomially bounded derivatives for which the degree of the
polynomial bound increases linearly with the order of the
derivative.

\section{How large are the extended Moyal algebras?}
The next theorem establishes inclusion relations between the duals
of the Palamodov spaces~\eqref{2.5} and~\eqref{2.5*} and the
algebras $\mathcal M_\theta(S_\alpha^\beta)$ and $\mathcal
C_\theta(S^\alpha_\beta)$.

{\bf Theorem~6:} {\it Let $\alpha\ge \beta$. For any $\theta$, the
space
$C'(S_\alpha^\beta)=\bigcup\limits_{B\to\infty}\bigcap\limits_{A\to
\infty}E_{\alpha,A}^{\beta,B}$ is contained in the algebra
$\mathcal M_\theta(S_\alpha^\beta)$ and the space
$M'(S^\alpha_\beta)=\bigcup\limits_{B\to\infty}\bigcap\limits_{A\to
\infty}(E^{\alpha,A}_{\beta,B})'$ is contained in the algebra
$\mathcal C_\theta(S^\alpha_\beta)$.

Proof.}  In the theorem's formulation, we use the presentation of
$C'(S_\alpha^\beta)$ and $M'(S^\alpha_\beta)$ proved in Theorem~2.
Let $u\in C'(S_\alpha^\beta)$. We need show that for any $f\in
S_\alpha^\beta$, the product  $u\star f$ also belongs to
$S_\alpha^\beta$. By definition~\eqref{3.1},
\begin{equation}
\langle u\star f,h\rangle=\langle u,f\star h\rangle \label{6.3}
 \end{equation}
for all $h\in S_\alpha^\beta$.  We let $L_f$ denote the linear map
$w\to f\star w$ from $(S_\alpha^\beta)'$ into $C(S_\alpha^\beta)$.
By Corollary of Theorem~5, this map is continuous. Consequently,
its transpose $L'_f$ is well defined as a map from
$C'(S_\alpha^\beta)$ into $(S_\alpha^\beta)^{\prime\prime}$. The
second dual  coincides with $S_\alpha^\beta$ because this space is
reflexive,  hence $L'_fu\in S_\alpha^\beta$. For all $w\in
(S_\alpha^\beta)'$, we have the equality
\begin{equation}
\langle w, L'_f u\rangle=\langle u,f\star w\rangle.
 \label{6.4}
 \end{equation}
If $w=h\in S_\alpha^\beta$, then the right-hand side
of~\eqref{6.4} becomes equal to that of~\eqref{6.3}  and the
left-hand side of~\eqref{6.4} takes the form $\int (L'_f
u)(x)h(x)dx$. Thus, the function $L'_f u$ considered as an element
of $(S_\alpha^\beta)'$ coincides with the functional $u\star f$
and we conclude that $u\in {\mathcal
M}_{\theta,L}(S_\alpha^\beta)$. Analogously, $u\in {\mathcal
M}_{\theta,R}(S_\alpha^\beta)$ and hence $u\in {\mathcal
M}_\theta(S_\alpha^\beta)$. Now let $v\in M'(S^\alpha_\beta)$.
Then $\mathcal F^{-1}v\in C'(S_\alpha^\beta)$ by Theorem~1 and
$\mathcal F^{-1}v\in {\mathcal M}_\theta(S_\alpha^\beta)$ by what
has just been said. Therefore, $v$ belongs to $\widehat{{\mathcal
M}_\theta(S_\alpha^\beta)}={\mathcal C}_\theta(S^\alpha_\beta)$,
which completes the proof.

Of special interest are the Moyal multiplier algebras of the
Fourier-invariant spaces $S^\beta_\beta$.

{\bf Theorem~7:} {\it Suppose that $\beta\ge1/2$ and the matrix
$\theta$ is invertible. In this case, the Moyal multiplier algebra
$\mathcal M_\theta(S^\beta_\beta)$ contains both the spaces
$C'(S_\beta^\beta)$ and $M'(S_\beta^\beta)$. The algebra $\mathcal
C_\theta (S^\beta_\beta)$ also contains these spaces and consists
of the same elements as $\mathcal
M_{-4\theta^{-1}}(S^\beta_\beta)$.

Proof.} Changing variables in one of integrals~\eqref{1.2}, we
obtain
\begin{equation}
(f\star_\theta g)(x)= \frac{1}{\pi^d\det\theta}\int  f(x-\xi) \hat
g(2\theta^{-1}\xi)\,e^{2ix\theta^{-1}\xi}d\xi=\frac{1}{\pi^d\det\theta}
(f\mathbin{\hat{\star}}_{-4\theta^{-1}}\mathcal F_\theta g)(x),
\notag
\end{equation}
where $(\mathcal F_\theta g)(\xi)=\int
g(x)e^{-2ix\theta^{-1}\xi}dx$ is the symplectic Fourier transform
of $g$, which  obviously belongs to $S^\beta_\beta$ if $g\in
S^\beta_\beta$. An analogous manipulation with another integral
in~\eqref{1.2} gives
\begin{equation}
(f\star_\theta g)(x)=\frac{1}{\pi^d\det\theta}(\overline{\mathcal
F}_\theta f\mathbin{\hat{\star}}_{-4\theta^{-1}}g)(x),
 \label{6.5}
\end{equation}
where $(\overline{\mathcal F}_\theta f)(\xi)=\int
f(x)e^{2ix\theta^{-1}\xi}dx$.  Because $S^\beta_\beta$ is dense in
$(S^\beta_\beta)'$, the Moyal product and the twisted convolution
product have  unique continuous extensions to the case where one
of factors is in $(S^\beta_\beta)'$, and we conclude that
\begin{equation}
u\star_\theta g=\frac{1}{\pi^d\det\theta} u\hstar_{-4\theta^{-1}}
\mathcal F_\theta g\quad \text{and}\quad f\star_\theta u=
\frac{1}{\pi^d\det\theta} \overline{\mathcal F}_\theta
f\hstar_{-4\theta^{-1}}u
 \label{6.6}
\end{equation}
for each $u\in(S^\beta_\beta)'$ and for all $f,g\in
S^\beta_\beta$. Therefore, an element of $(S^\beta_\beta)'$
belongs to $\mathcal M_\theta(S^\beta_\beta)$ if and only if it
belongs to $\mathcal C_{-4\theta^{-1}}(S^\beta_\beta)$.

Combining Theorem~5 with~\eqref{6.6}, we see that  the products
$u\star_\theta g$ and $f\star_\theta u$ are contained in
$M(S^\beta_\beta)$. Moreover, the maps $(g,u)\to u\star_\theta g$
and $(f,u)\to f\star_\theta u$ are hypocontinuous from
$S^\beta_\beta\times (S^\beta_\beta)'$ into $M(S^\beta_\beta)$.
The rest of proof is similar to the proof of Theorem~6. Let $v\in
M'(S_\beta^\beta)$. We   show that for  any $g\in S_\beta^\beta$,
the product $g\star_\theta v$ also belongs to $S_\beta^\beta$.
By~\eqref{3.1}, we have
\begin{equation}
\langle g\star_\theta v,h\rangle=\langle v,h\star_\theta g\rangle
 \label{6.8}
 \end{equation}
for all $h\in S_\beta^\beta$. Let $R_g$ be the continuous linear
map $u\to u\star_\theta g$ from $(S_\beta^\beta)'$ into
$M(S_\beta^\beta)$. Then
\begin{equation}
\langle u, R'_g v\rangle=\langle v,u\star_\theta g\rangle
 \label{6.9}
 \end{equation}
for all $u\in (S_\beta^\beta)'$. If $u=h\in S_\beta^\beta$, then
the right hand sides of~\eqref{6.8} and \eqref{6.9} coincide and
$\langle h, R'_g v\rangle=\int  (R'_g v)(x)h(x)dx$. Since $h$ is
an arbitrary element of $S_\beta^\beta$, we conclude that
$g\star_\theta v=R'_g v$ and hence $v\in {\mathcal
M}_{\theta,R}(S_\alpha^\beta)$. Analogously, $v\in {\mathcal
M}_{\theta,L}(S_\beta^\beta)$ and consequently $v\in {\mathcal
M}_\theta(S_\beta^\beta)$. This result together with the
equalities $\widehat{M(S_\beta^\beta)}=C(S^\beta_\beta)$ and
$\widehat{{\mathcal M}_\theta(S_\beta^\beta)}={\mathcal
C}_\theta(S^\beta_\beta)$ implies that $C'(S^\beta_\beta)\subset
\mathcal C_\theta(S^\beta_\beta)$. Thus, Theorem~7  is proved.

{\it Remark:} Theorem~7 shows that the algebra ${\mathcal
M}_\theta(S_\beta^\beta)$ is invariant under the Fourier transform
if (and really only if) $\theta^2/4=-I$. In the phase space
representation of quantum mechanics, one usually uses  the
symplectic Fourier transform, which is natural when dealing with
the Weyl correspondence. In this connection, it should be noted
that either of the two operators  ${\mathcal F}_\theta$ or
$\overline{\mathcal F}_\theta$ maps $\mathcal
M_\theta(S^\beta_\beta)$ isomorphically onto $\mathcal
C_{-4\theta^{-1}}(S^\beta_\beta)$, and these algebras consist of
the same elements.

Like the Moyal multiplier algebra of the Schwartz space, any
algebra $\mathcal M_\theta(S^\beta_\beta)$ with $\beta\ge1/2$
contains all polynomials and all distributions of compact support.
Moreover, Theorem~6 shows that for $\beta>1$,  this algebra
contains all smooth functions that belong to the  Gervey
class~\cite{H1} of order $\beta$ and  grow at infinity not faster
than exponentially of order $1/\beta$, type 0. By Theorem~7, it
also contains all  ultradistributions of the Roumieu~\cite{R,Kom}
class ${\{\on^{\beta \on}\}}$  that decrease at infinity not
slower than exponentially of order $1/\beta$, finite type. The
elements of $(S^1_1)'$ are called
Fourier-hyperfunctions~\cite{Kaw}, and the algebra $\mathcal
M_\theta(S^1_1)$ contains all real-analytic functions growing not
faster than exponentially of order 1, type 0 and also all
Fourier-hyperfunctions decreasing not slower than order~1, finite
type. Analogous statements hold for the analytic functionals
defined on the spaces $S^\beta_\beta$ with $\beta<1$.

\section{A special role of the algebra $\mathcal
M_\theta(S^{1/2}_{1/2})$}

As noted in Sec.~II, the space $S^\beta_\beta$ is nontrivial if
and only if $\beta\ge1/2$. The algebra $\mathcal
M_\theta(S^{1/2}_{1/2})$ plays a special role in the theory. By
Theorem~6, it contains the space
\begin{equation}
E^{1/2}_{1/2}=\projlim_{A\to\infty,B\to0} E^{1/2,B}_{1/2,A},
\label{7.1}
 \end{equation}
which is a linear subspace of $C'(S^{1/2}_{1/2})$. By the
arguments used in  the proof of Theorem~2, $E^{1/2}_{1/2}$ is an
FS space. Using Taylor's formula and Cauchy's inequality, it is
easy to verify that $E^{1/2}_{1/2}(\oR^d)$ coincides with the
space of restrictions to $\oR^d$ of all entire functions on
$\oC^d$ that are of  order at most two, type zero. In other words,
these entire functions satisfy the inequalities $|f(z)|\leq
C_{f,\epsilon}\, e^{\epsilon|z|^2}$, where $\epsilon>0$ and can be
taken arbitrarily small. It is readily seen that
series~\eqref{1.1} converges pointwise for such functions.
Moreover, the following theorem holds.

{\bf Theorem 8:} {\it The space $E^{1/2}_{1/2}$ is  a topological
algebra with respect to the Moyal star product. If $f,g \in
E^{1/2}_{1/2}$, then  series~\eqref{1.1} representing
$f\star_\theta g$ converges absolutely in the topology of this
space.

Proof.}  Let $G(s)=\sum_\on c_\on s^\on$  be an entire function of
order  $<2$ or of order 2 and finite type. We will show that then
the differential operator $G(\partial)$ is a continuous
endomorphism of $E^{1/2}_{1/2}$ and the series $\sum_\on
c_\on\partial^\on f$, where $f\in E^{1/2}_{1/2}$, converges
absolutely in every norm of this space. According to the theory of
entire functions (see, e.g., Sec.~IV.5.2 of~\cite{GS2}) the above
bound on the growth of $G(s)$ is equivalent to the condition
\begin{equation}
|c_\on|\leq C(b/\on)^{\on/2},
 \label{7.2}
\end{equation}
where $C$ and $b$ are positive constants. The space
$E^{1/2}_{1/2}$ consists of all functions such that
\begin{equation}
|\partial^\on f(x)|\le \|f \|_{-A,
B}{B^{|\on|}\on^{\on/2}}\,e^{|x/A|^2},
 \label{7.3}
\end{equation}
for any $A,B>0$. Let $B'\ge B\sqrt{2}$. Using~\eqref{7.2},
\eqref{7.3}, and the inequality $(n+m)^{n+m}\le 2^{|n+m|}n^nm^m$,
we obtain
\begin{equation}
\|c_n\partial^\on f\|_{-A,B'}=\sup_{x,\om}\left|\,
\frac{c_\on\partial^{\on+\om}
f(x)}{B^{\prime|\om|}\om^{\om/2}}\,e^{-|x/A|^2}\right|\le C\|f
\|_{-A, B}(B\sqrt{2b})^{|\on|}.
 \label{7.4}
\end{equation}
Because $B$ can be taken arbitrarily small, we conclude that the
series $\sum_\on c_\on\partial^\on f$ converges absolutely in
every norm of $E^{1/2}_{1/2}$. Moreover, the factor $\|f \|_{-A,
B}$ on the right-hand side of~\eqref{7.4} shows that the map
$E^{1/2}_{1/2}\to E^{1/2}_{1/2}\colon f\to G(\partial)f$ is
continuous. In particular, the operator
$e^{\frac{i}{2}\theta^{ij}\partial_{x^i}\partial_{y_j}}$ is well
defined and acts continuously on $E^{1/2}_{1/2}(\oR^{2d})$. The
Moyal product $f\star_\theta g$ is obtained by applying this
operator to the function $(f\otimes g)(x,y)$ in
$E^{1/2}_{1/2}(\oR^{2d})$ and then identifying  $x$ with $y$. It
is easily verified that the restriction to the diagonal $x=y$ is a
continuous map from $E^{1/2}_{1/2}(\oR^{2d})$ into
$E^{1/2}_{1/2}(\oR^d)$, which completes the proof.

It should be emphasized that $E^{1/2}_{1/2}$ is the largest star
product algebra with the property of absolute convergence of the
series determining this product. Indeed, the bound $|\partial^\on
f(0)|\le C_BB^{|\on|}\on^{\on/2}$, $\forall B>0$, on the
derivatives of $f$ at the origin implies that $f$ cannot grow
faster than with order 2 and minimum type. This subalgebra of
$\mathcal M_\theta(S^{1/2}_{1/2})$ in turn has various subalgebras
which are specified by additional restrictions on the behavior of
their elements at the infinity of  real space. In particular,
$E^{1/2}_{1/2}$ contains the  space $\mS^{1/2}=\projlim_{B\to0,
N\to\infty}S^{1/2,B}_N$, where $S_N^{1/2,B}$ is the Banach space
of  analytic functions such that
\begin{equation}
\|f \|_{B,N}\defeq\sup_{x,\on}\,(1+|x|)^N\frac{|\partial^\on
f(x)|}{B^{|\on|}\on^{\on/2}}<\infty.
  \label{7.5}
\end{equation}
As shown in~\cite{S07}, the space $\mS^{1/2}$ is  a topological
algebra with respect to the Moyal star product and, being adequate
to the nonlocal nature of this product, is suitable for using as a
test function space  in a general formulation of quantum field
theory on noncommutative space-time. The definitions~\eqref{1.1}
and \eqref{1.2} of Moyal multiplication are equivalent for
functions in this space and the algebras $(S,\star)$ and
$(E^{1/2}_{1/2},\star)$ can be regarded as different extensions of
$(\mS^{1/2},\star)$. Moreover, to each closed cone $V\subset
\oR^d$ we can assign an algebra $\mS^{1/2}(V)$ which also consists
of entire functions and is defined similarly but with supremum
over $x\in V$ in an analog of~\eqref{7.5}. All the algebras
$\mS^{1/2}(V)$ are also contained in $E^{1/2}_{1/2}$. If a
functional $u\in (\mS^{1/2})'$ has a continuous extension to
$\mS^{1/2}(V)$, then the cone $V$ can be thought of as a carrier
of $u$.  As argued in~\cite{S07, S10}, the spaces $S_N^{1/2,B}(V)$
can be used as a tool for formulating causality in noncommutative
quantum field theory. Another family of subalgebras of
$E^{1/2}_{1/2}$ is formed by the spaces
$\mS^{1/2}_\alpha=\projlim_{A\to\infty,B\to0}S^{1/2,B}_{\alpha,A}$,
 $\alpha>1/2$, and by their siblings associated with cones in
$\oR^d$.  The basic reason for considering the spaces over cones
is explained in~\cite{S95}. It lies in the fact that the
continuous functionals defined on the $S$-type spaces with
superscript $\beta<1$ retain the angular localizability property,
in spite of the failure of the notion of support in the case of
entire analytic test functions.

\section{Concluding remarks}
In this paper, we content ourselves with considering the Moyal
multiplier algebras of the spaces $S^\beta_\alpha$. However the
general construction of Sec.~III is applicable to any test
function space on which the Weyl-Heisenberg group acts
continuously and whose topological properties are more or less
like those of $S$. In particular, it  immediately extends to the
Gel'fand-Shilov spaces $S^b_a$ and $W^\Omega_M$ specified by more
flexible restrictions on the smoothness and  behavior at infinity
of their elements. (See Supplements~1 and 2 in~\cite{GS2} for the
definition of these spaces.)

The Schwartz space $S$ can formally be considered as a limit of
the spaces $S^\beta_\beta$ as $\beta\to\infty$, and an analog of
Lemma~1 shows the existence of an approximation of the identity
for $S$. Namely, if  $f\in S$, $e\in S$, and $e(0)=1$, then  the
sequences of operators of left and right Moyal multiplication by
$e(x/\nu)$ converges to the unit operator in the topology of
${\mathcal L}(S)$ as $\nu\to \infty$. Because of this, an analog
of Theorem~1 also holds for $S$, which gives an alternative
definition of the algebras ${\mathcal M}_{\theta,L}(S)$ and
${\mathcal M}_{\theta,R}(S)$, different from the original
definition~\cite{A1,A2,G-BV1,G-BV2}. It should be mentioned that
the existence of an approximation of the identity with the weaker
property of pointwise convergence on elements of $S$ was
previously indicated in~\cite{Mail}.

It is worth noting that the Fr\'echet space  $E^{1/2}_{1/2}$ is a
topological algebra not only with respect to the Moyal product but
also with respect to the Wick star product. (We refer the reader
to~\cite{BS} for the definition and main properties of this
product which also is often called the Wick-Voros product.) The
proof of this fact is similar to that of Theorem~8, with the
replacement of
$e^{\frac{i}{2}\theta^{ij}\partial_{x^i}\partial_{y_j}}$ by the
bi-differential operator corresponding to the Wick product.
Thereby we obtain a simple and explicit solution to the
problem~\cite{BRW} of constructing the largest Fr\'echet space of
analytic functions for which the Wick star product converges and
depends continuously on the deformation  parameter. As shown
in~\cite{S10} the space $\mS^{1/2}$  is also a topological algebra
with respect to the Wick product. Finally we note that the Weyl
transformation can be naturally extended to the Moyal multipliers
discussed here and their definition can be expressed in terms of
the corresponding operators on a Hilbert space, but this is beyond
the scope of this paper.
\medskip

\medskip

\section*{Acknowledgments}
 This paper was supported in part
by the the Russian Foundation for Basic Research (Grant
No.~09-01-00835) and the Program for Supporting Leading Scientific
Schools (Grant No.~LSS-1615.2008).

\section*{Appendix: Proof of Lemma~2}
For each function $f\in S(\oR^d)$ and for any multiindices $\ok$,
$\on$, the following inequality holds:
$$
 \int_{\oR^d}|\partial^\ok
 x^\on|\,|f(x)|\,dx\le\sqrt{2}\int_{\oR^d}|x^\on|\,
 |\partial^\ok f(x)|\,dx,
 \eqno{({\rm A}1)}
$$
It suffices to prove (A1) for functions of one variable and for
$\ok=1$, because the general case can be easily reduced to this
one. We suppose first that  $f(x)$ is a real-valued function on
$\oR$ and divide the semiaxis $x\ge0$ into the three parts $M^+$,
$M^-$, and $M^0$, where $f(x)$ takes positive, negative, and zero
values respectively. After numbering the connected components of
$M^+$ and $M^-$, we  write
$$
 \int_0^\infty
 (x^n)'|f(x)|\,dx=\sum_j\int_{M^+_j}f(x)dx^n
 -\sum_j\int_{M^-_j}f(x)dx^n\le \int_0^\infty x^n|f'(x)|dx.
  $$
Thus, (A1) holds for all real-valued functions in $S(\oR)$ even
without the coefficient $\sqrt{2}$. This coefficient is relevant
to the case of complex-valued functions because
$|u|+|v|\le\sqrt{2}|u+iv|$.

Let $f\in S^{\beta, B}_{\alpha,A}$ and $A'>d^\alpha A$. Using (A1)
and the inequality $\sum_{j=1}^d |x_j|^{1/\alpha}\le d\,
|x|^{1/\alpha}$, we get
\begin{multline}
|p^\om\partial^\on \hat f(p)|=\left|\int
e^{-ipx}\partial^\om[x^\on f(x)]dx\right|\le
\int\sum_\ok\binom{\om}{\ok}|\partial^\ok
x^\on|\,|\partial^{\om-\ok}f(x)|dx\le \\\le \sqrt{2} \int
2^{|\om|} |x^\on
\partial^\om f(x)|dx\le C_{A'}\|f\|_{A,B}(2B)^{|\om|}
\om^{\beta\om}\sup_x |x^\on|\prod_j e^{-|x_j/A'|^{1/\alpha}}.
\notag
\end{multline}
The supremum over  $x$ on the right-hand side equals
$A^{\prime|\on|}(\alpha/e)^{\alpha|\on|}\on^{\alpha\on}$ and hence
\begin{multline}
|\partial^\on \hat f(p)|\le
C_{A'}\|f\|_{A,B}A^{\prime|\on|}(\alpha/e)^{\alpha|\on|}\on^{\alpha\on}
\inf_\om\frac{(2B)^{|\om|} \om^{\beta\om}}{|p^\om|}\\\le C
\|f\|_{A,B}A^{\prime|\on|}(\alpha/e)^{\alpha|\on|}\on^{\alpha\on}
e^{-(\beta/e)|p/2B|^{1/\beta}}. \notag
\end{multline}
A subtlety in the calculation of infimum over $\om$ is  that the
components of  $\om$ ranges over integers, but this affects only
the magnitude of the coefficient  $C$. We conclude
that~\eqref{5.1} holds for any $r>\max\{(\alpha d/e)^\alpha,
2(e/\beta)^\beta\}$. Lemma~2 is proved.


\baselineskip=13pt

\begin{thebibliography}{99}

\bibitem{I}  M.~A.~Soloviev, {\it Star product algebras of test functions},
Theor. Math. Phys. {\bf 153}  1351-1363  [arXiv:0708.0811].

\bibitem{GS2} I.~M.~Gelfand and G.~E.~Shilov, {\it Generalized
Functions},  Vol.~2, Academic,  New York, 1964.

\bibitem{BS}  F.~A.~Berezin and M.~A.~Shubin, {\it Schr\"odinger
equation,} Kluwer, Dordrecht, 1991.


\bibitem{Sz} R.~J.~Szabo, {\it Quantum field theory on noncommutative
spaces},  Phys. Rep. {\bf 378} (2003) 207-299
[arXiv:hep-th/0109162].

\bibitem{DFR} S.~Doplicher, K.~Fredenhagen, and J.~E.~Roberts,
{\it The quantum structure of space-time at the Planck scale and
quantum fields},  Commun. Math. Phys. {\bf 172} (1995) 187-220
[arXiv:hep-th/0303037].

\bibitem{SWitten}  N.~Seiberg and E.~Witten, {\it String theory and
noncommutative geometry},  JHEP {\bf 9909} (1999) 032
[arXiv:hep-th/9908142].

\bibitem{Gr} H.~J.~Groenewold, {\it On the principles of elementary quantum
mechanics},  Physica {\bf 12} (1946)  405-460.

\bibitem{Moyal} J.~E.~Moyal, {\it  Quantum mechanics as a statistical
theory},  Proc. Cambridge Phil. Soc. {\bf 45} (1949)  99-124.

\bibitem{A1} M.~A.~Antonets, {\it The classical limit for Weyl quantization},
Lett. Math. Phys. {\bf 2} (1978)  241-245.

\bibitem{A2}M.~A.~Antonets, {\it  Classical limit of Weyl
quantization},  Theor. Math. Phys. {\bf 38} (1979) 219-228.

\bibitem{Kamm} J-B.~Kammerer, Analysis of the Moyal product in a flat
space,  {\it J. Math. Phys.} \textbf{27} (1986)  529-535.

\bibitem{Mail} J.~M.~Maillard, {\it On the twisted convolution
product and the Weyl transformation of tempered distributions}, J.
Geom. Phys. {\bf 3} (1986) 230-261.

\bibitem{G-BV1} J.~M.~Gracia-Bondia and J.~C.~V\'arilly, {\it Algebras
of distributions suitable for phase-space quantum mechanics. I},
J. Math. Phys. \textbf{29} (1988)  869-879.

\bibitem{G-BV2} J.~C.~Varilly and J.~M.~Gracia-Bondia, {\it Algebras
of distributions suitable for phase-space quantum mechanics. II.
Topologies on the Moyal algebra},  J. Math. Phys. \textbf{29}
(1988) 880-887.

\bibitem{E} R.~Estrada, J.~M.~Gracia-Bondia, and J.~C.~V\'arilly,
{\it On asymptotic expansions of twisted products}, J. Math. Phys.
\textbf{30} (1988) 2789-2796.

\bibitem{G-BV3} J.~M.~Gracia-Bondia, F.~lizzi, G.~Marmo, and
P.~Vitale, {\it Infinitely many star products to play with}, JHEP
{\bf 0204} (2002) 026  [arXiv:hep-th/0112092].


\bibitem{G} V.~Gayral, J.~M.~Gracia-Bondia, B.~Iochum, T.~Sch\"ucker,
 and J.~C.~V\'arilly, {\it Moyal planes are spectral triplets},
Commun. Math. Phys. \textbf{246} (2004) 569-623
[arXiv:hep-th/0307241].

\bibitem{Alv}  L.~Alvarez-Gaume and M.~A.~Vazquez-Mozo,
{\it General properties of noncommutative field theories},  Nucl.
Phys. B {\bf 668} (2003) 293-321 [arXiv:hep-th/0305093].

\bibitem{Ch} M.~Chaichian, M.~N.~Mnatsakanova,  A.~Tureanu,
and Yu.~A.~Vernov, {\it Test function space in  noncommutative
quantum field theory},  JHEP {\bf 0809} (2008) 125
[arXiv:0706.1712].

\bibitem{Green} O.~W.~Greenberg, {\it Failure of microcausality in quantum
field theory on noncommutative spacetime},  Phys.Rev.  D {\bf 73}
(2006)  045014 [arXiv:hep-th/0508057].

\bibitem{S08} M.~A.~Soloviev, {\it On the failure of microcausality in
noncommutative field theories},  Phys. Rev. D {\bf 77} (2008)
125013 [arXiv:0802.0997].

\bibitem{SW} R.~F.~Streater and A.~S.~Wightman,
{\it PCT, Spin and Statistics and All That,} Benjamin, New York,
1964.

\bibitem{BLOT}  N.~N.~Bogoliubov, A.~A.~Logunov, A.~I.~Oksak,
 and I.~T.~Todorov, {\it General Principles of Quantum Field Theory,}
 Kluwer, Dordrecht, 1990.

\bibitem{S07} M.~A.~Soloviev, {\it  Noncommutativity and $\theta$-locality},
J. Phys A: Math. Theor. {\bf 40} (2007) 14593-14604
[arXiv:0708.1151].

\bibitem{S10} M.~A.~Soloviev, {\it Noncommutative deformations of quantum
field theories, locality and causality}, Theor. Math. Phys. {\bf
163} (2010) 741-752 [arXiv:1012.3536].

\bibitem{S06} M.~A.~Soloviev, {\it Axiomatic formulations of
nonlocal and noncommutative field theories}, Theor. Math. Phys.
{\bf 147} (2006)  660-669  [arXiv:hep-th/0605249].

\bibitem{FS1} A.~Fisher, R.~J~Szabo, {\it Duality covariant quantum
field theory on noncommutative Minkowski space}, JHEP {\bf 0902}
(2009) 031 [arXiv:0810.1195].

\bibitem{FS2} A.~Fisher, R.~J~Szabo, {\it  UV/IR duality in noncommutative
quantum field theory},  Gen. Relativ. Gravit. {\bf 43} (2010)
2509-2522 [arXiv:1001.3776].

\bibitem{Zahn} J.~Zahn, {\it Divergences in quantum field theory on the
noncommutative two-dimensional   Minkowski space with
Grosse-Wulkenhaar potential},  Ann. Henri Poincar\'e {\bf 12}
(2011) 777-804 [arXiv:1005.0541].

\bibitem{MV} R.Meise and D.Vogt, {\it Introduction to Functional
Analysis,} Clarendon, Oxford, 1997.

\bibitem{M} B.~S.~Mityagin, {\it Nuclearity and other properties of spaces of type
S}, Amer. Math. Soc. Transl., Ser. 2, {\bf 93}, Amer. Math. Soc.,
Providence, RI (1970), pp.~45-59.

\bibitem{K}  G.~K\"othe, {\it Topological Vector Spaces II,} Springer,
New York, 1979.

\bibitem{P} V.~P.~Palamodov, {\it  Fourier transforms of rapidly
increasing infinitely differentiable functions},  Trudy Moskov.
Mat. Obshch. {\bf 11} (1962) 309-350 [in Russian].

\bibitem{Sch}  H.~H.~Schaefer, {\it Topological Vector Spaces,}
MacMillan, New York,  1966.

\bibitem{S82} M.~A.~Solov'ev, {\it Spacelike asymptotic behavior
of vacuum expectation values in nonlocal  field theory},  Theor.
Math. Phys. {\bf 52} (1982)  854-862.

\bibitem{H1} L.~H\"ormander, {\it The Analysis of Linear Partial
Differential Operators I. Distribution Theory and Fourier
Analysis}, Springer-Verlag, Berlin, 1983.

\bibitem{R} C.~Roumieu, {\it Sur quelques extensions de la notion de
distribution},  Ann. Sci. \'Ecole Norm., Sup\'er. (3), {\bf 77}
(1960) 41-121.

\bibitem{Kom} H.~Komatsu, {\it Ultradistributions, I. Structuure theorems
and a characterization},  J. Fac. Sci. Univ. Tokyo, Sec. IA, {\bf
20} (1973) 25-105.

\bibitem{Kaw} T.~Kawai, {\it On the theory of Fourier hyperfunctions
and its applications to partial differential equations with
constant coefficients},  J. Fac. Sci. Univ. Tokyo, Sec. IA, {\bf
17} (1970) 467-517.

\bibitem{S95} M.~A.~Soloviev, {\it Towards a generalized distribution
 formalism for gauge quantum fields},  Lett. Math. Phys. {\bf 33}
(1995) 49-59   [arXiv:hep-th/9403083].

\bibitem{BRW} S.~Beiser, H.~R\"omer, S.~Waldman, {\it  Convergence
of the Wick star product},  Commun. Math. Phys. {\bf 272} (2007)
25-52 [arXiv:math.QA/0506605].




\end{thebibliography}
\end{document}